\documentclass{jfm}
\usepackage{graphicx}
\usepackage{pdflscape}
\usepackage{epstopdf,epsfig}
\usepackage{newtxtext}
\usepackage{newtxmath}
\usepackage{natbib}
\usepackage{multirow}
\usepackage{hyperref}
\usepackage{booktabs}
\usepackage[dvipsnames]{xcolor}
\usepackage{rotating}
\usepackage{pdflscape}
\definecolor{myblue}{RGB}{0, 0, 230}
\hypersetup{
    colorlinks = true,
    linkcolor  = myblue,
    urlcolor   = myblue,
    citecolor  = myblue,
}

\useAMSsubequations
\newcommand{\RomanNumeralCaps}[1]
\linenumbers

\usepackage{lineno}

\shorttitle{Supersonic jet impingement on concave surfaces}
\shortauthor{H. Chandravamsi, D. V. Shenoy and S. H. Frankel}

\title{Supersonic jet impingement on concave surfaces}

\author{
Hemanth Chandravamsi\aff{1,2}\corresp{\email{hemanth@campus.technion.ac.il}},
Dhanush Vittal Shenoy\aff{1}
\and
Steven H. Frankel\aff{1}
}

\affiliation{
\aff{1} Faculty of Mechanical Engineering, Technion - Israel Institute of Technology,
Haifa 3200003, Israel
\aff{2} Department of Mechanical Engineering, Johns Hopkins University,
Baltimore, MD 21218, USA
}

\begin{document}

\raggedbottom

\date{\today}

\maketitle

\begin{abstract}
The aeroacoustic resonance of round supersonic jets impinging on concave surfaces is investigated using compressible large-eddy simulations, vortex-sheet modelling, and Powell's feedback-loop analysis. The choked jets operate at an ideally expanded Mach number of $1.56$ and a Reynolds number of $6\times10^4$. Six geometries are considered: two flat plates at $L/D=2.08$ and $2.58$, where $L$ is the nozzle-to-wall distance and $D$ the nozzle exit diameter, and four Gaussian concave surfaces of fixed depth and indentation spread $\sigma\in\{0.4,0.8,1.6,4.0\}$. As the indentation narrows, the primary-tone amplitude increases by up to $23\,\mathrm{dB}$ relative to the flat-wall reference at $L/D=2.6$, together with larger wall-pressure fluctuations and moments. A Powell-Tam source-transfer budget attributes this amplification to increased Mach-disk source amplitude and more efficient return of the upstream feedback wave to the nozzle. The stronger upstream-propagating waves are consistent with acoustic focusing by the concave wall. For the helical cases, the measured frequencies and radial eigenfunctions agree closely with the guided jet mode predicted by the vortex-sheet model, supporting its role in closing the upstream feedback path. The same selection is recovered for concave and flat walls alike, so this tone is governed by the shear-layer profile of the equivalent ideally expanded jet rather than by the wall geometry. The axisymmetric frequencies, by contrast, coincide with no guided-mode branch and appear instead to follow Powell's classical loop-length criterion. The results identify distinct frequency-selection mechanisms for helical and axisymmetric screech and demonstrate that wall curvature provides effective control of screech amplitude and surface loading.



\end{abstract}


\section{Introduction} \label{sec:intro}
The inherent unsteady nature of the flow field of non-ideally expanded supersonic jets, and the noise they generate, have been subjects of interest since the seminal work of Alan Powell \citep{powell1953mechanism}. The noise field in a wide range of non-ideally expanded supersonic jet configurations is dominated by a distinct high-amplitude tonal component known as screech \citep{edgington2019aeroacoustic}, typically falling within the range 2 kHz to 15 kHz, with near-field sound pressure levels that can exceed 150 dB. The aeroacoustic resonance mechanism that guides this tone production was first theorised by \citet{powell1953mechanism} through his classical feedback loop model. According to this theory, hydrodynamic instability waves are generated near the nozzle lip, and convect downstream along the jet shear layer to interact with the shock cells. The waves undergo strong amplification upon interaction, creating a local coherent acoustic source (sometimes an array of sources). Subsequently, the generated acoustic waves travel upstream towards the nozzle lip region, closing the feedback loop and thus sustaining the generation of initial instability waves. 

The presence of an impinging surface normal to the jet axis amplifies the tonal intensity of the radiated acoustic field \citep{wagner1971penetration, back1978pressure}. Under sufficiently close nozzle to plate distances, it was reported \citep{neuwerth1973thesis, ho1981dynamics} that unlike a mid-column interaction between instability waves and the jet, the downstream propagating waves tend to amplify near the Mach-disk region in impinging jets (edge tones). Based on this observation, \citet{ho1981dynamics} and \citet{nosseir1982dynamics} have extended the feedback loop model of free jets by considering the downstream and upstream wave travel lengths as the impingement distance between nozzle exit and the plate. The feedback loop mechanism and the consequent staging behavior in supersonic jets was found to be in good agreement with the experimental observations of \citet{wagner1971penetration} and \citet{umeda1987discrete}, among others. Although the downstream propagating waves from the nozzle to plate are well-established to be guided by the Kelvin-Helmholtz instability waves (of the shear layer), the mechanism behind upstream propagating waves was not fully resolved. Support from both externally propagating acoustic waves which travel exterior to the jet \citep{ho1981dynamics,umeda1987discrete} and the internal waves that propagate through the jet column \citep{wagner1971penetration,neuwerth1974acoustic} have been debated. Addressing this question, \citet{tam1990theoretical} proposed a vortex-sheet model in support of the internal, jet-column route, identifying the upstream-propagating waves that close the feedback loop as the jet's `lowest-order intrinsic neutral modes'. Dispersion relations were derived to predict the wavenumber of these modes for varying jet Mach numbers and nozzle-to-plate spacings. For subsonic jets, these upstream-propagating neutral waves are found to be confined primarily inside the jet. Recent high-fidelity simulations indicate that in ideally expanded supersonic impinging jets, the upstream-propagating component of the feedback loop is governed by guided jet modes (GJMs) confined within the jet plume \citep{bogey2017feedback,maia2024tones,wilke2016origin}. For free screeching jets, the same GJM closes the loop and predicts the tone frequencies more accurately than an externally propagating acoustic wave \citep{gojon2018oscillation,gojon2019antisymmetric,mancinelli2019screech}. In under- or over-expanded jets with strong shock interactions, the feedback may instead be sustained through external acoustic or shock-leakage waves \citep{edgington2021generation}.

This distinction is consequential for the present study. A GJM is confined inside the plume, and its dispersion is fixed by the shear-layer profile; the frequency it selects is therefore largely insensitive to the external shape of the impinging surface, which enters primarily through the loop length \citep{tam1990theoretical,bogey2017feedback}. An externally propagating acoustic wave, by contrast, radiates through the ambient, where the concave wall can redirect it toward the nozzle in the manner of a concave mirror. Which of the two pathways dominates for non-ideally expanded impinging jets remains unsettled. The relative importance of the internal and external contributions depends on the shock strength and the nozzle-to-plate standoff, and the two pathways may coexist. The present configuration, at $M_j=1.56$, lies squarely in this regime. A focusing of the upstream leg by the concave wall, through the external acoustic pathway, is therefore plausible but not assured \textit{a priori}, and is one of the questions investigated in the present study.


Adding to the predictions of the theoretical models, various experimental and numerical studies have investigated the screech tones and feedback waves of impinging supersonic jets. In the case of ideally expanded jets, \citet{krothapalli1999flow}, \citet{elavarasan2000piv} have experimentally investigated the flow, lift, and noise characteristics of round impinging jets using Particle Image Velocimetry (PIV) and acoustic measurements. Later on, consistent with the theory of \citet{tam1990theoretical}, \citet{gojon2016investigation}, \citet{bogey2017feedback} have numerically demonstrated and characterised the existence of upwind propagating waves that close the aeroacoustic feedback loop for various nozzle to plane distances. In the context of non-ideally expanded supersonic jets, experimental investigations such as by \citet{powell1988sound}, \citet{henderson1993experiments}, and \citet{henderson2005experimental} have addressed physical aspects concerning the tone production, mode staging, and the effect of nozzle to wall distance. \citet{henderson1993experiments} note that the Mach disk plays an important role in sustaining the feedback loop in non-ideally expanded jets. At some nozzle-to-plate distances, where the Mach disk is not close to the plate, this mechanism is absent. This is further supported by the experimental observations of \citet{akamine2014experimental}. \citet{mitchell2012visualization} have employed an ultra high-speed schlieren visualisation setup to directly observe for the first time the downstream and upstream propagating waves of the feedback loop responsible for the screech tones in the impinging jets. Complementing these experiments, \citet{uzun2013simulation} have performed large-eddy simulations (LES) of a Mach 1.5 impinging jet to examine the effect of jet temperature and the links between dominant coherent structures and the primary screech tone. \citet{brehm2016noise} have numerically investigated the location of noise source in a supersonic jet impinging on inclined flat plate using two point cross-correlation analysis. \citet{houston2018simulations} have combined simulations and experiments of dual high-speed impinging jets, and \citet{zigunov2019instability} have characterised the instability modes of millimetre-scale supersonic jets experimentally. \citet{li2023acoustic} measured the axisymmetric screech modes of underexpanded jets impinging on an inclined plate over $M_j$ from $1.05$ to $1.56$, and showed that the feedback loops are closed by upstream-propagating GJMs whose selected frequencies shift as the plate alters the shock-cell spacing, demonstrating that an external boundary can modulate the resonance. More recently, the numerical study of \citet{gojon2017flow} uncovered various oscillation modes, flow unsteadiness, and the linked tone intermittency at the same jet operating condition considered here. 

Beyond the radiated tones, non-ideally expanded impinging jets under specific operating conditions also display various forms of symmetric and asymmetric flow structure oscillations within the jet column. \citet{ginzburg1970some} observed large-scale Mach-disk oscillations with amplitudes comparable to the nozzle diameter, accompanied by a pulsating recirculation bubble near the impingement location. The experimental studies of \citet{nakatogawa1971disintegration} and \citet{semiletenko1974features} note that the presence and the relative position of the Mach disk plays an important role in the flow structure instability. More recently, a similar phenomenon was observed by \citet{sinibaldi2015sound} under high nozzle-pressure-ratio conditions, where strong Mach disk oscillations were noted to influence the interaction between the shear-layer and the plate. The combined experimental and numerical study of \citet{sakakibara2002oscillation} has presented axisymmetric, helical, and non-regular jet oscillation modes occurring at different impinging wall distances. The PIV study of \citet{henderson2005experimental} provides a detailed investigation of Mach disk unsteadiness, the flow within the recirculation bubble, and the attached peripheral planar flow over the plate. They point out that the presence of flow structure oscillations isn't necessarily guaranteed by the creation of a recirculation zone; instead, it might predominantly depend on the strength of the Mach disk. \citet{risborg2009high} have documented eight distinct flow structure instability modes in impinging jets through the use of ultra high-speed schlieren and shadowgraph visualisation. They systematically varied the nozzle pressure ratio, impingement distance, and impinging plate angle to observe different instability modes. They also detail pulsing, flapping, and helical shock motions within the jet plume, along with a periodic appearance and decay of the wall Mach disk in some cases. Such flow unsteadiness, especially near the impinging wall, can lead to `fatigue loading' which can be catastrophic. However, literature addressing the loading characteristics of impinging jets is sparse, and is examined in the present study.

The oscillations catalogued in these studies are conventionally classified as symmetric or asymmetric oscillation modes \citep{powell1992observations,norum1983screech}. The occurrence of these modes is linked to the geometrical nature of the incoming vortical structures, or more precisely, the azimuthal phase instabilities that surround the jet. Vortical structures with toroidal, spiral, and a combination of two counter-rotating spiral shaped instabilities produce oscillation modes namely `toroidal', `helical', and `flapping' respectively \citep{tam1990theoretical}. While analytical solutions are possible for all of these vortical configurations, the most stable mode, or combination of modes, depends primarily on the nozzle pressure ratio and the impingement distance \citep{ponton1997near}. The selected mode has been observed to be the one with the largest screech amplification, or with the most efficient transmission of shear-layer instability and acoustic waves within the feedback loop.

Although there are a number of reported studies on supersonic jets impinging on inclined flat-plate surfaces (jet-blast-deflector configurations), research on curved impinging walls in the literature is quite sparse. \citet{jennions1980axisymmetric} have performed experiments on supersonic jets impinging on a co-axially placed conical surface. They note a wide variety of shock patterns arising due to the interaction of the cone shock with the jet shock and briefly comment on the shock structure unsteadiness. \citet{prasad1994impingement} have performed experiments and Euler computations to study the loading characteristics on the axisymmetric jet deflector wall. However, their study mainly focused on the mean flow quantities and mean pressure exerted on the deflector wall. More recently, \citet{mason2015shock} have performed experiments to study the effect of impinging wall curvature on flow and acoustics using concave and convex cylindrical sections as impinging walls. They note that a concave cylindrical surface can potentially suppress the impingement tone formation by deflecting the wall jet flow into the entrainment field and thus creating a similar effect as that of nozzle tabs used to control the jet noise \citep{reeder1996evolution}. Geometry is one of several routes to tone control; active flow control has also been used to reduce the noise of cold and hot supersonic jets \citep{zigunov2022reduction}. A related class of configurations is that in which an underexpanded jet is directed normally into a cavity, tube or cup and sustains intense self-excited tones. This is the basis of the Hartmann whistle \citep{hartmann1922new} and the Hartmann-Sprenger resonance tube \citep{brocher1970fluid,murugappan2005parametric,kastner2002development,raman2009powered}, in which the tone amplitude and frequency are set by the cavity depth and the jet-to-cavity standoff. A concave impinging surface presents a comparable cavity to the jet, so the wall curvature may influence the screech response through an analogous cavity coupling in addition to its effect on the feedback path.

The small number of reported studies on concave impinging walls leaves the underlying interaction between the jet and the curved wall poorly understood, and the link between wall geometry, mode selection and screech-tone frequency remains unclear. The present study considers supersonic jets impinging on concave walls, represented by Gaussian indentations of fixed depth and differing spread $\sigma$, alongside two flat-wall references, and examines their flow unsteadiness, surface loading and associated screech tones. The near-field acoustic fluctuations are related to the unsteady flow field, and the dominant jet oscillation mode is identified for each configuration. The observed tone frequencies are then compared with both Powell's feedback-loop estimate and the vortex-sheet guided-mode dispersion relation, allowing the roles of global loop length and upstream-wave dispersion to be distinguished. The wall shape is expected to modify the impingement-region flow, alter the effective feedback pathway, and thereby change the flow-acoustic coupling, radiated directivity, and unsteady wall loading. Such configurations arise in practical settings, including vertical take-off and landing over uneven or contoured ground \citep{mehta2013thruster}, jet-blast deflectors \citep{tsutsumi2014acoustic}, cooling of contoured turbine blades, and cold-spray deposition over curved surfaces \citep{mahdavi2018analytical}.


The paper is organised as follows. The computational approach, the jet parameters, and the impinging-wall geometry are described in \S\ref{sec:setup}. The instantaneous and mean flow fields, the near-field pressure spectra, the screech-tone modes, the unsteady Mach-disk motion, and a source-transfer budget for the curvature-dependent tonal amplification are presented in \S\ref{sec:flowstructure}, together with comparison against experiments and previous LES. The unsteady forces and moments on the impinging surface are quantified in \S\ref{sec:loading}. The downstream convection velocity and the structure of the upstream-propagating waves are extracted from the LES and compared with the vortex-sheet dispersion relation in \S\ref{sec:waves}. These are assembled into feedback-loop and guided-jet-mode predictions of the tone frequencies in \S\ref{sec:feedback}, and concluding remarks are given in \S\ref{sec:conclusion}.

\section{Jet setup and numerical methodology} \label{sec:setup}

\subsection{Computational setup of the jet}
\begin{figure}
    \centering
    \includegraphics[width=\textwidth]{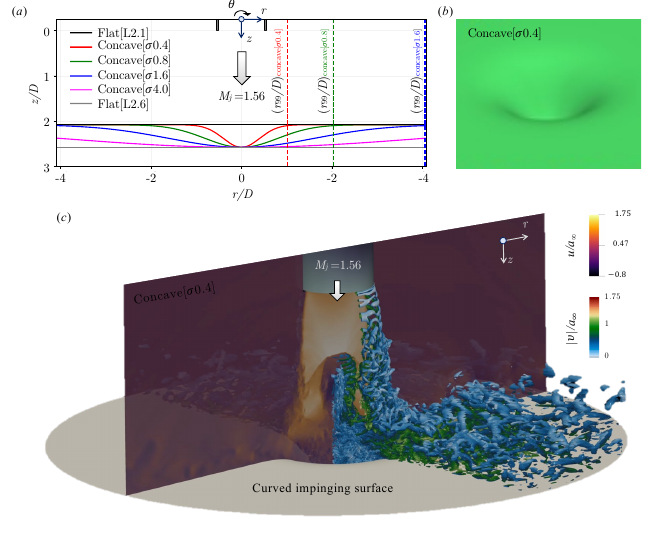}
    \caption{(a) Geometric profiles of the flat and concave impinging surfaces defined by equation~(\ref{eqn:profile}). (b) Isometric view of the Concave[$\sigma$0.4] surface illustrating the localised indentation. (c) Instantaneous flow field of the Concave[$\sigma$0.4] jet, showing the axial velocity $u/a_\infty$ in the meridional $r$-$z$ plane and iso-surfaces of the Q-criterion coloured by the velocity magnitude $|\mathbf{u}|/a_\infty$.}
    \label{fig:jetWalls}
\end{figure}

The present simulations replicate the impinging-jet configuration investigated experimentally by \citet{henderson2005experimental} and adopt the computational setup of \citet{gojon2017flow}. The computational domain, shown in figure~\ref{fig:jetWalls}(a), consists of a circular jet-exit plane of diameter $D$ with a lip thickness of $0.05D$ and an impinging surface located downstream. The internal nozzle flow is not included in the computational domain. Six impinging configurations are examined: two flat and four concave surfaces. The first flat-wall case employs a nozzle-to-plate spacing of $L=2.08D$, consistent with the reference experiments \citep{henderson2005experimental} and LES studies \citep{gojon2017flow}, and serves as the validation case. For the concave configurations lying between the two flat-wall limits, the impinging-surface geometry is parameterised by a Gaussian profile,
\begin{equation} \label{eqn:profile}
    \frac{z}{D}= 2.08 + 0.5 \, \exp{\left[-\frac{(r/D)^2}{2\sigma^2}\right]},
\end{equation}
where $r$ is the radial distance from the jet axis. The parameter $\sigma$ controls the radial decay of the wall indentation. The limiting cases $\sigma \to 0$ and $\sigma \to \infty$ correspond to the flat-wall configurations Flat[L2.1] and Flat[L2.6], respectively. The influence of the indentation spread $\sigma$ on the flow structure oscillations and near-field acoustics is examined within these bounds. Two derived nozzle-to-wall distances are referenced below: the axial wall position along the lip line $r/D=0.5$,
\begin{equation} \label{eqn:ell_lip}
    \ell_{\rm lip}/D = 2.08 + 0.5\,\exp(-0.125/\sigma^2),
\end{equation}
and the area-weighted mean wall position over the jet-column footprint $r\le 0.5\,D$,
\begin{equation} \label{eqn:L_avg}
    L_{\rm avg}/D = L/D + 4\sigma^2\,[1 - \exp(-0.125/\sigma^2)],
\end{equation}
both of which reduce to $L/D$ in the flat-wall limit. The computational domain extends to $r/D=40$ radially and $z/D=-20$ upstream of the jet exit plane.

\begin{table}
\begin{center}
\def~{\hphantom{0}}
\begin{tabular}{lcccccc}
Case name & $L/D$ & $\sigma$ & Curvature depth & $r_{99}/{D}$ & $\ell_{\rm lip}/{D}$  & $L_{\rm avg}/D$  \\[4pt]
Flat[L2.1]           & 2.08 &  0 (limit) & - & - & 2.08   & 2.08    \\
Concave[$\sigma$0.4] & - &  0.4          & 0.5$D$  & 1.21  & 2.31  & 2.43       \\
Concave[$\sigma$0.8] & - &  0.8          & 0.5$D$  & 2.43  & 2.49  & 2.53       \\
Concave[$\sigma$1.6] & - &  1.6          & 0.5$D$  & 4.86  & 2.56  & 2.57        \\
Concave[$\sigma$4.0] & - &  4.0          & 0.5$D$  & 12.14 & 2.58 & 2.58       \\
Flat[L2.6]           & 2.58 &  $\infty$ (limit) & -  & -  & 2.58 & 2.58   \\
\end{tabular}
\caption{Summary of geometric parameters for simulated jets. $L/D$ is the non-dimensional nozzle-to-surface distance, and $\sigma$ is the spread (standard deviation) of the Gaussian indentation equation~(\ref{eqn:profile}). The radial coordinate $r_{99}/D = \sigma\sqrt{2\ln 100} \approx 3.03\,\sigma$ marks where the indentation has decayed to $1\%$ of its depth; it quantifies the radial extent of the indentation. The column $\ell_{\rm lip}/D$ is the axial distance from the nozzle exit to the impinging surface measured along the lip line $r/D = 0.5$ equation~(\ref{eqn:ell_lip}). The last column $L_{\rm avg}/D$ is the area-weighted mean axial wall position over the jet-column footprint $r \le 0.5\,D$ equation~(\ref{eqn:L_avg}).}
\label{tab:cases}
\end{center}
\end{table}

\noindent \textit{Jet parameters.} The jet exhaust conditions are sonic and identical across all configurations, defined using the isentropic relations:
\begin{subequations} \label{eqn:exit_conditions}
\begin{gather}
\frac{p^{\text{exit}}}{\rho_{\infty}a_{\infty}^2}
=\frac{1}{\gamma}\left[\frac{2+(\gamma-1)M_j^2}{\gamma+1}\right]^{\frac{\gamma}{\gamma-1}}, \quad
\frac{\rho^{\text{exit}}}{\rho_{\infty}}
=\frac{\gamma(\gamma+1)}{2(T_0/T_{\infty})}
\left(\frac{p^{\text{exit}}}{\rho_{\infty}a_{\infty}^2}\right),
\tag{\theparentequation a,b}\\[4pt]
\frac{u_z^{\text{exit}}}{a_{\infty}}
=\sqrt{\frac{2(T_0/T_{\infty})}{\gamma+1}}, \quad
u_r^{\text{exit}}=u_\theta^{\text{exit}}=0.
\tag{\theparentequation c,d}
\end{gather}
\end{subequations}
The ideally expanded jet Mach number is set to $M_j=1.56$, consistent with the experiments of \citet{henderson2005experimental}, corresponding to a nozzle pressure ratio $P_0/P_{\infty}=4.03$. The total-to-ambient temperature ratio is set to $T_0/T_{\infty}=1.0$. Following \citet{gojon2017flow}, the jet Reynolds number is set to $\text{Re}_j = u_j D_j / \nu_j = 6\times10^4$, where $D_j$, $u_j$ (439.7 m/s) and $\nu_j$ are the diameter, velocity and kinematic viscosity of the ideally expanded equivalent jet. This smaller Reynolds number, relative to the experiments, is chosen to allow adequate resolution of the turbulent structures while maintaining computational feasibility. A smooth, unperturbed boundary-layer profile is prescribed at the jet exit. Employing a boundary-layer thickness of $\delta = 0.05D$, the axial velocity in the region $0.5D-\delta \le r < 0.5D$ is specified as $u_z = u_z^{\text{exit}}\,[(0.5D-r)/\delta]^{1/7}$, with $u_z^{\text{exit}}$ the sonic exit velocity given by equation~(\ref{eqn:exit_conditions}), so that the profile decays from the core value to zero at the nozzle lip. The corresponding density variation is determined from the Crocco-Busemann relation. Frequencies are reported throughout as the jet-based Strouhal number $St = f\,D_j/u_j$, formed with the ideally-expanded jet diameter $D_j$ and velocity $u_j$.

\noindent \textit{Grid parameters.} A structured multi-block mesh comprising four blocks and approximately $127\times10^6$ cells is employed for all the jet configurations. The singularity along the jet axis, arising from the axisymmetric geometry, is avoided by using a butterfly mesh topology as described in \citet{chandravamsi2023application}. The grid refinement and spacings are based on the configuration reported by \citet{gojon2017flow}. The azimuthal direction is discretised using 512 uniformly distributed cells. In the axial direction, the grid is clustered near the nozzle exit and the impinging surface, maintaining a minimum spacing of $\Delta z = 0.00375D$ and a maximum spacing of $\Delta z = 0.08D$, with a total of 420 cells between the jet-exit-plane and the impinging surface. In the radial direction, the grid is refined near $r=0.5D$, where the minimum spacing is $\Delta r = 0.002D$. The spacing gradually increases to $\Delta r = 0.03D$ at $r=2.5D$, using 148 grid cells in the range $0.5D \le r \le 2.5D$, adequately resolving waves up to $St = 5.3$. Beyond $r=2.5D$, the mesh spacing is stretched from $\Delta r = 0.03D$ to $\Delta r = 0.04D$ within $2.5D \le r \le 7.5D$. The grid resolution in terms of LES wall units $\Delta r^{+}$, $(r\Delta\theta)^{+}$, and $\Delta z^{+}$ are identical to that of the configuration in \citet{gojon2017flow}. 


Sponge layers \citep{bodony2006analysis} combined with characteristic outflow conditions \citep{poinsot1992boundary} were applied at all far-field boundaries to limit reflections entering the computational domain. Calculations were performed until a non-dimensional time of $t a_{\infty}/D = 910$. After the transient phase, statistics were collected from $t a_{\infty}/D = 500$ using a uniform sampling interval of $\Delta t a_{\infty}/D = 0.05$, which provided 8192 snapshots for each case.


\subsection{Numerical methodology}
The LES approach used in this study follows the methodology established and validated in the authors' earlier works \citep{chandravamsi2024high, gichon2024dynamics, kakumani2024impinging, heppner2026shock}. Three-dimensional unsteady compressible Navier-Stokes equations are solved in generalised curvilinear coordinates. The equations are formulated in conservative non-dimensional form using the ambient reference quantities: temperature $T_{\infty}=293~\mathrm{K}$, speed of sound $a_{\infty}=343~\mathrm{m/s}$, density $\rho_{\infty}=1.205~\mathrm{kg/m^3}$, and molecular viscosity $\mu_{\infty}=1.83\times10^{-5}~\mathrm{Pa\cdot s}$. The viscous stresses are modeled using Newton's law of viscosity under Stokes' hypothesis with a bulk viscosity of zero. Temperature-dependent dynamic viscosity $\mu$ is determined from Sutherland's law. The fluid is treated as calorically perfect with a specific heat ratio of $\gamma=1.4$ and Prandtl number $\text{Pr}=0.71$.

The governing equations are solved using an in-house solver \citep{chandravamsiAPSGFM2023} that employs an optimised low-dispersion finite-difference approach. The inviscid fluxes are discretised with the sixth-order Optimized Upwind Reconstruction Scheme (OURS6) of \citet{chandravamsi2024high}, which uses a nine-point stencil. Regions of shocks are resolved using the modified monotonicity-preserving (MP) limiter of \citet{ahn2020modified}. Conservative fluxes are computed using a hybrid HLL-HLLC (Harten-Lax-van Leer with contact) approximate Riemann solver, where a Ducros sensor \citep{ducros1999large} selects the HLL scheme in shock regions and the HLLC elsewhere to mitigate the carbuncle phenomenon. The viscous fluxes are evaluated using the sixth-order midpoint-based formulation (ME6-Opti) of \citet{chandravamsi2024high}. The inherent numerical dissipation of the approximate Riemann solver and the numerical scheme is employed to dissipate the subgrid-scale energy, consistent with the implicit LES approach of \citet{ahn2021numerical}. Temporal integration is performed using the third-order total variation diminishing (TVD) Runge-Kutta scheme of \citet{gottlieb1998total}. All computations are executed on four NVIDIA A100 (40GB) GPUs using a parallelised code, and each configuration required approximately 18 days of compute time.

\begin{landscape}
\begin{figure}
    \centering
    \includegraphics[width=0.9\linewidth]{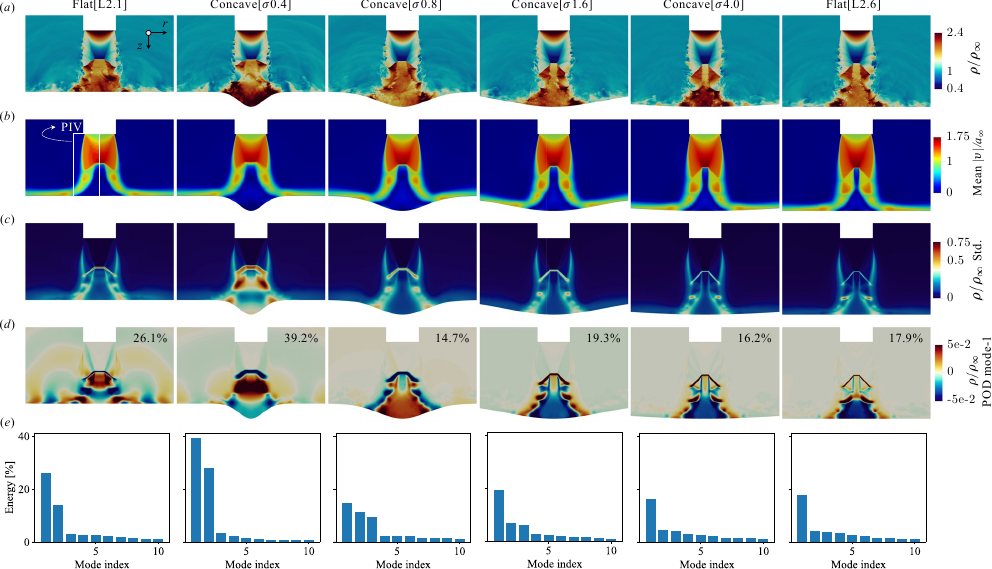}
    \caption{Instantaneous fields, statistics and modal content for the six impinging jet configurations.
    (a) Instantaneous density field.
    (b) Mean velocity magnitude field (LES); the white box in the Flat[L2.1] panel insets the experimental PIV of \citet{henderson2005experimental} over $-0.95\le r/D\le0$ for side-by-side comparison, all other fields being LES.
    (c) Standard deviation of the density field.
    (d) First POD mode based on the density time series with corresponding modal energy indicated.
    (e) Energy distribution of the first ten POD modes.}
    \label{fig:collage}
\end{figure}
\end{landscape}

\section{Flow-structure oscillations and screech tones} \label{sec:flowstructure}
\subsection{Instantaneous and mean flow field} \label{sec:flowfield}

The instantaneous and mean flow fields in figure~\ref{fig:collage} show that the upstream development of the jet plume is largely unaffected by the downstream wall geometry, with the influence of wall curvature confined primarily to the impingement region. All configurations exhibit the characteristic shock-cell train downstream of the nozzle exit, a quasi-normal Mach disk upstream of the impinging surface, and a radially deflected wall jet. Over the first $\approx 1.5D$ downstream of the nozzle exit, the shear-layer development and the shock-cell pattern are closely similar across all six cases. The Mach disk is the exception. It both moves downstream and contracts as the indentation is widened (\S\ref{sec:machstem}), so the wall geometry does reach the terminal shock, while leaving the shear layer that feeds it unchanged. The largest differences are in the impingement region, where the indentation spread controls the size of the recirculation zone and the subsonic pocket between the Mach disk and the impinging surface (figure~\ref{fig:collage}b). The instantaneous $Q$-criterion iso-surfaces in figure~\ref{fig:jetWalls}(c) reveal two distinct turbulent shear layers within the plume. The outer shear layer develops from the nozzle lip and separates the jet from the ambient fluid, while an internal shear layer originates at the triple point of the Mach reflection and separates the subsonic recirculating flow behind the Mach disk from the surrounding supersonic stream. Between these two shear layers, an annular supersonic jet forms, containing a sequence of trapped expansion and compression waves that undergo repeated reflections between the inner and outer shear layers.

The mean velocity magnitude field, $\overline{|\mathbf{\upsilon}|}/a_{\infty}$, and the standard deviation of the density fluctuation, are shown in figure~\ref{fig:collage}(b,c). The LES reproduces the experimental PIV measurements of \citet{henderson2005experimental} for the Flat[L2.1] configuration with good qualitative agreement, capturing the jet core, the recirculation region downstream of the Mach disk, and the annular jet. The influence of wall curvature is most evident downstream of the Mach disk. The extent of the recirculation region first increases from Flat[L2.1] to Concave[$\sigma$0.8], before decreasing toward that of the Flat[L2.6] configuration. This non-monotonic variation indicates that the redistribution of the incoming axial momentum into the wall-parallel direction depends sensitively on the wall curvature, reaching a maximum at intermediate indentation spreads. The possible bearing of this enlarged recirculation region on the tone amplification is taken up in \S\ref{sec:nearfield}. The fluctuation field exhibits three regions of high variance: the shear layer separating the supersonic core from the entrained ambient fluid, the Mach disk, and the annular jet between the inner and outer shear layers. The fluctuation amplitude in the Mach-disk region is largest for Concave[$\sigma$0.4], where the density standard deviation approaches $0.75\rho_{\infty}$, followed by Flat[L2.1] and Concave[$\sigma$0.8]; for the three least-curved configurations, the corresponding peak value is reduced by approximately a factor of two. In contrast, the fluctuation levels within the outer shear layer remain comparable across all configurations over the first shock cell.

The dominant coherent content of the unsteady density field is examined using proper orthogonal decomposition (POD) \citep{weiss2019tutorial}. The leading POD mode is shown in figure~\ref{fig:collage}(d), and the corresponding energy distribution of the first ten modes is presented in figure~\ref{fig:collage}(e). The leading mode captures $26.1\%$ of the total fluctuation energy for Flat[L2.1] and $39.2\%$ for Concave[$\sigma$0.4], whereas the corresponding values for the remaining four configurations lie between $14.7\%$ and $19.3\%$. The comparatively large energy of the leading mode in Flat[L2.1] and Concave[$\sigma$0.4] indicates the dominance of a single coherent oscillation, consistent with the sharply tonal near-field acoustic spectra reported in \S\ref{sec:nearfield}. The spatial structure of the leading mode separates the same two groups. Flat[L2.1] and Concave[$\sigma$0.4] exhibit a coherent axisymmetric structure centred on the Mach disk and the impingement region, consistent with a toroidal jet-column oscillation. In the remaining four configurations no single structure dominates. Their POD energy spectra in figure~\ref{fig:collage}(e) decay more gradually, with the energy distributed over the first three or four modes, consistent with the pairwise eigenvalue near-degeneracy associated with rotating travelling-wave modes \citep{edgington2019aeroacoustic}. This grouping matches the azimuthal-mode classification obtained from the exit-plane phase fields in \S\ref{sec:modes}, where Flat[L2.1] and Concave[$\sigma$0.4] are axisymmetric and the other four configurations helical.

\subsection{Near-field pressure spectra and curvature-dependent tone amplification} \label{sec:nearfield}

\begin{figure}
    \centering
    \includegraphics[width=\textwidth]{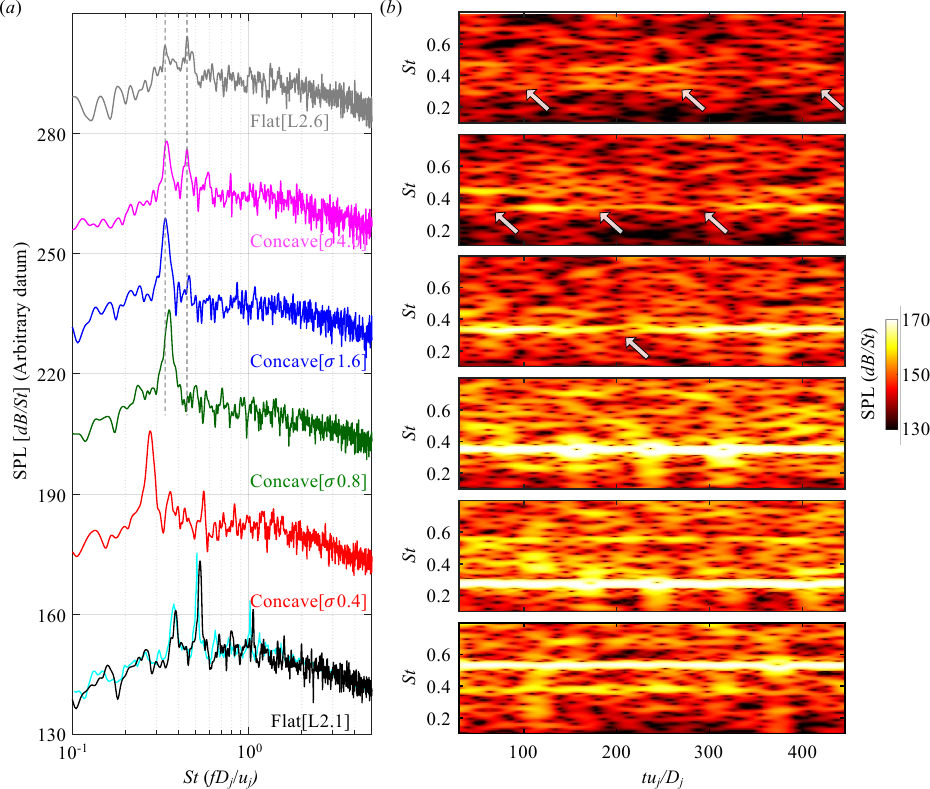}
    \caption{
        (a) Near-field pressure spectra at $r/D=1$ for the flat and concave configurations. The cyan curve is the Flat[L2.1] case from the LES of \citet{gojon2017flow}.
        (b) Spectrograms of the same signals. The rows follow the same top-to-bottom order as the offset spectra in panel~(a), from Flat[L2.6] at the top to Flat[L2.1] at the bottom. Arrows indicate intermittent breaks or decreased amplitude in tonal events.
    }
    \label{fig:fft_probes}
\end{figure}

The near-field pressure spectra are obtained from a uniformly-spaced azimuthal ring of pressure probes on the nozzle-exit plane at $r/D=1$. The power spectral density (PSD) of the azimuthally-averaged probe signal, expressed in dB/$St$ relative to the reference pressure $p_{\text{ref}}=2\times10^{-5}~\mathrm{Pa}$, is presented in figure~\ref{fig:fft_probes}(a) for the six configurations, with the spectra vertically offset by +30~dB for clarity. The Flat[L2.1] spectrum (black) is overlaid with the LES result of \citet{gojon2017flow} (cyan), obtained at nominally identical jet operating conditions. The broadband levels agree closely, the median difference between the two spectra being below $1$~dB across the resolved Strouhal range. Both show the same three tonal features: a dominant tone, its first harmonic, and a weaker secondary peak. These lie at $St=0.529$, $1.060$ and $0.382$ in the present LES, and at $St=0.510$, $1.018$ and $0.377$ in that of \citet{gojon2017flow}, so the reference tones fall $1$-$4\%$ lower. Their peak levels agree to within $3.5$~dB.

Figure~\ref{fig:fft_probes}(a) and table~\ref{tab:tones} show that the impinging-wall geometry affects both the dominant-tone frequency and amplitude. The tone shifts from $St_1=0.529$ for Flat[L2.1] to $St_1=0.275$ for Concave[$\sigma$0.4], while the remaining concave cases cluster near $St_1\approx 0.34$-$0.36$. The two narrowest indentations, Concave[$\sigma$0.4] and Concave[$\sigma$0.8], produce the highest primary-tone levels, which decrease as the spread increases towards the flat-wall limit. The peaks for the three widest-spread cases, Concave[$\sigma$1.6], Concave[$\sigma$4.0] and Flat[L2.6], are mutually close in both frequency and amplitude, so the spectrum approaches the flat-wall result once the indentation extends well beyond the jet column. Both trends appear to be governed by the radial reach of the indentation relative to the jet column: the strongest tones arise for the configurations whose indentation remains radially confined ($r_{99}/D\lesssim 2.4$, table~\ref{tab:cases}), whereas a broad indentation reaching well into the entrained ambient leaves the spectrum close to that of a flat wall. The curved wall would then act on the resonance primarily where its indentation overlaps the jet column and the impingement region, rather than through its far-field geometry. The azimuthal character of these tones is established in \S\ref{sec:modes}.

\begin{table}
\begin{center}
\def~{\hphantom{0}}
\begin{tabular}{lccccc}
Case  & $St_1$ (dominant) & $St_2$ & $St_3$ & $\mathrm{SPL}_1$ (dB/$St$) & OASPL (dB) \\[4pt]
Flat[L2.1]           & 0.529 (A) & 0.382 (C) & 1.060 (A) & 173.1 & 158.0  \\
Concave[$\sigma$0.4] & 0.275 (A) & 0.365 (C) & 0.552 (A) & 175.7 & 160.5  \\
Concave[$\sigma$0.8] & 0.357 (C) & 0.706 (C) & -         & 175.3 & 160.6  \\
Concave[$\sigma$1.6] & 0.336 (C) & 0.439 (C) & -         & 167.8 & 154.8  \\
Concave[$\sigma$4.0] & 0.342 (C) & 0.449 (C) & -         & 157.1 & 150.1  \\
Flat[L2.6]           & 0.348 (C) & 0.435 (C) & -         & 152.7 & 148.7  \\
\end{tabular}
\caption{Strouhal numbers of the screech tones, the primary-tone sound pressure level $\mathrm{SPL}_1$, and the azimuthally-averaged overall sound pressure level (OASPL), measured on the nozzle-exit-plane probe ring at $r/D=1$. Axisymmetric and helical modes are labelled `A' and `C' respectively.}
\label{tab:tones}
\end{center}
\end{table}

Time-frequency representations of the pressure signal, obtained via short-time Fourier transform with a Hann-windowed segment of $\Delta t u_j/D_j = 25$ and $75\%$ overlap, are shown in figure~\ref{fig:fft_probes}(b). The dominant tones of Flat[L2.1] and Concave[$\sigma$0.4] appear as continuous horizontal bands of nearly uniform intensity, indicating that the screech feedback loop is sustained throughout the observation window. Concave[$\sigma$0.8] also carries an unbroken band of comparable strength, though its amplitude varies along the record. The three widest-spread configurations, Concave[$\sigma$1.6], Concave[$\sigma$4.0] and Flat[L2.6], instead show repeated breaks and amplitude drops in the tonal band, indicated by arrows. In these cases the tonal energy reorganises between adjacent Strouhal bands over time intervals of $tu_j/D_j \sim 50$-$100$. Such intermittency has been reported in impinging and screeching supersonic jets and is commonly associated with mode staging between competing feedback-loop states \citep{henderson1993experiments, raman1999supersonic, edgington2015staging, gojon2017flow, bell2021intermittent, leon2022three, lee2023super, li2023acoustic}. Across the present six cases the continuity of the tonal band orders with its amplitude: the three configurations with the highest $\mathrm{SPL}_1$ (table~\ref{tab:tones}) sustain an unbroken band, whereas the three with the lowest sustain the loop only intermittently.

\begin{figure}
    \centering
    \includegraphics[width=\textwidth]{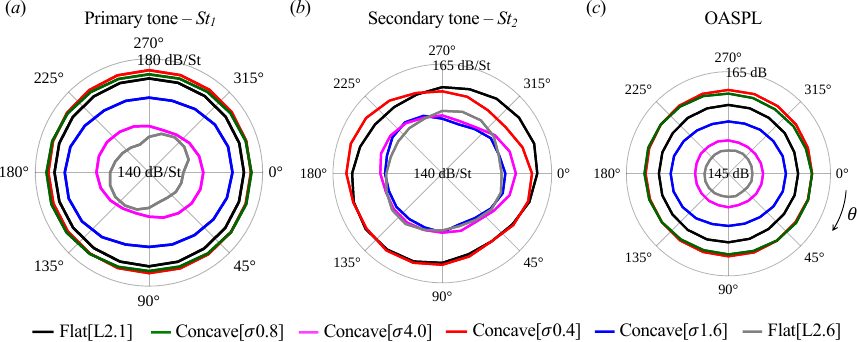}
    \caption{Polar distributions of sound pressure level on the nozzle-exit plane at $r/D=1$: band-pass-filtered power at (a) the primary screech tone $St_1$ and (b) the secondary tone $St_2$, and (c) the overall level (OASPL). The acoustic loading increases monotonically across the concave family as the indentation spread $\sigma$ decreases, the largest primary-tone amplification arising for the narrowest indentation.}
    \label{fig:SPL}
\end{figure}

The azimuthal distribution of the acoustic loading on the nozzle-exit-plane probe ring is presented in figure~\ref{fig:SPL}. Figure~\ref{fig:SPL}(a) and (b) report the band-pass-filtered SPL at the primary and secondary screech tones, respectively, while figure~\ref{fig:SPL}(c) reports the overall sound pressure level (OASPL) obtained by integrating the PSD over the full band resolved by the sampling interval, up to the Nyquist limit $St = 8.6$. For all configurations the polar distributions are nearly axisymmetric, with maximum deviations from the azimuthally-averaged value below $\pm 1.5$~dB. This uniformity is expected: it follows from the long-time averaging of the rotating pressure pattern in the asymmetric (helical or flapping) cases, and from the intrinsic axisymmetry of the symmetric (toroidal) cases. The radial extent of the polar curves orders the cases by acoustic intensity: the primary-tone SPL increases monotonically across the concave family as the indentation spread $\sigma$ decreases, from $\approx 153$~dB/$St$ for Flat[L2.6] to $\approx 176$~dB/$St$ for Concave[$\sigma$0.4], an amplification of $\approx 23$~dB that exceeds the corresponding shift in OASPL. The flat-wall limit $\sigma\to0$ does not continue this trend, the Flat[L2.1] primary tone lying $2.6$~dB below that of Concave[$\sigma$0.4] at a shorter standoff. Thus reducing the indentation spread acts not only to redistribute the acoustic energy but also to concentrate a larger fraction of it in the tonal component of the spectrum. While the dominant tone shifts in frequency and changes its azimuthal symmetry as the indentation narrows, much as it does between the two flat plates as their spacing is increased, the distinctive effect of the wall curvature is the amplification of the tone. The two most strongly amplified cases, Concave[$\sigma$0.4] and Concave[$\sigma$0.8], lie within the intermediate-spread range over which the recirculation pocket behind the Mach disk is enlarged (\S\ref{sec:flowfield}). The concave indentation, into which the underexpanded jet is directed nearly normally, forms a wall cavity analogous to the resonator of a Hartmann whistle \citep{hartmann1922new,raman2009powered}, in which a jet impinging into a cavity radiates intense tones whose amplitude and frequency depend on the cavity size and the jet-to-cavity standoff. The enlarged cavity of the narrower indentations may therefore reinforce the resonance through a coupling of this kind, an effect that requires no particular azimuthal symmetry and so is compatible with both of these cases, whose oscillation modes differ (\S\ref{sec:modes}). The resulting amplification is quantified in the source-transfer budget of \S\ref{sec:why}.

\subsection{Mach-disk unsteadiness} \label{sec:machstem}

\begin{figure}
    \centering
    \includegraphics[width=\textwidth]{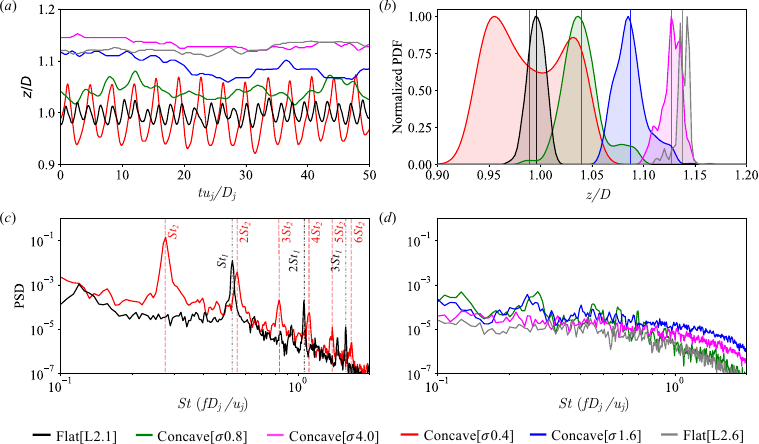}
    \caption{Mach-disk unsteadiness and associated spectral content.
    (a) Temporal evolution of the axial Mach-disk location along $r=0$.
    (b) Probability density function (PDF) of the Mach-disk location.
    (c) Power spectral density (PSD) of the Mach-disk motion for Flat[L2.1] and Concave[$\sigma$0.4]. Both spectra are dominated by a discrete fundamental at the near-field screech tone of the configuration, $St_1 = 0.529$ and $0.275$ respectively, followed by a harmonic sequence resolved to $3St_1$ and $6St_1$ above the broadband floor; the labelled harmonics identify the motion as a non-sinusoidal limit cycle rather than a narrow-band random oscillation.
    (d) PSD for the remaining four configurations, plotted on the same scale.
    }
    \label{fig:MachStem}
\end{figure}

The axial pulsation of the Mach disk provides the axisymmetric forcing for the feedback loop, but its coherence varies considerably across the six configurations. The instantaneous Mach-disk position on the jet axis, $z_{\text{ms}}(t)$, is tracked using the upstream-most location on $r=0$, and the resulting time histories are shown in figure~\ref{fig:MachStem}(a). Flat[L2.1] and Concave[$\sigma$0.4] exhibit highly regular periodic motion, with non-dimensional periods of $tu_j/D_j \approx 1.9$ and $3.6$, respectively. Concave[$\sigma$0.8] shows a weaker quasi-periodic response, whereas the remaining cases are dominated by irregular broadband motion. The overall displacement amplitude does not by itself distinguish these behaviours. Using the $1$st-to-$99$th-percentile range of $z_{\text{ms}}$ to exclude isolated tracking excursions gives $\Delta z_{\text{ms}}/D = 0.06$ for Flat[L2.1], $0.16$ for Concave[$\sigma$0.4], $0.10$ for Concave[$\sigma$0.8], $0.07$ for Concave[$\sigma$1.6], and $0.04$ for both Concave[$\sigma$4.0] and Flat[L2.6]. Thus, Concave[$\sigma$1.6] has a larger total displacement than the strongly periodic Flat[L2.1], despite lacking a coherent oscillation. The key distinction is therefore the tonal content: a band of half-width $\Delta St = 0.02$ about the primary tone retains $84\%$ and $93\%$ of the root-mean-square displacement for Flat[L2.1] and Concave[$\sigma$0.4], but only $5\%$ to $10\%$ for the other four cases.

The probability density functions of $z_{\text{ms}}$ in figure~\ref{fig:MachStem}(b) summarise the global statistics of the Mach-disk position. The mean location $\overline{z}_{\text{ms}}/D$ shifts monotonically downstream as the indentation spread $\sigma$ increases, from $0.99$ for Concave[$\sigma$0.4] to $1.14$ for Flat[L2.6], in agreement with the nozzle-to-plate spacing trend imposed by the wall indentation (table~\ref{tab:cases}). The distributions for Concave[$\sigma$0.4] (red) and Concave[$\sigma$0.8] (green) are markedly broader than those of the remaining configurations, with the Concave[$\sigma$0.4] distribution displaying a clearly bimodal shape consistent with a finite-amplitude limit-cycle oscillation. The distributions for the four least-tonal cases are narrowly peaked and approximately Gaussian, indicating a stochastic stationary motion of the Mach disk about its mean location.

The spectral content of the Mach-disk motion is presented in figure~\ref{fig:MachStem}(c) and (d). For Flat[L2.1] and Concave[$\sigma$0.4] (figure~\ref{fig:MachStem}c), the spectra are dominated by a fundamental tone followed by a sequence of higher harmonics. The fundamental Strouhal numbers $St_1=0.529$ for Flat[L2.1] and $St_1=0.275$ for Concave[$\sigma$0.4] coincide with the dominant near-field acoustic tones of these configurations (table~\ref{tab:tones}), and harmonics up to $3St_1$ and $6St_1$ respectively are resolved above the broadband floor. The presence of multiple harmonics is characteristic of a non-sinusoidal limit-cycle oscillation and is consistent with the asymmetric shape of the corresponding probability density functions in figure~\ref{fig:MachStem}(b). For the four remaining configurations (figure~\ref{fig:MachStem}d) the Mach-disk motion is broadband, with no clearly emerging discrete tone. This agrees with the predominantly helical jet-column oscillation reported for these cases in \S\ref{sec:flowfield}, since a rotating non-axisymmetric mode produces no net axial displacement of the Mach disk.

For the two axisymmetric configurations, the dominant tones of the Mach-disk motion coincide with those of the near-field pressure spectra (\S\ref{sec:nearfield}), showing that the axial pulsation of the Mach disk is phase-locked to the aeroacoustic feedback loop. The pulsation is then the principal source of normal-force unsteadiness on the impinging surface (\S\ref{sec:forces}). This is in line with the experimental observations of \citet{henderson2005experimental} and \citet{sinibaldi2015sound}, who attributed the strong axial Mach-disk motion to a resonant coupling between the impingement-region flow and the screech feedback loop.

\subsection{Characterisation of jet oscillation modes} \label{sec:modes}

\begin{figure}
    \centering
    \includegraphics[width=\textwidth]{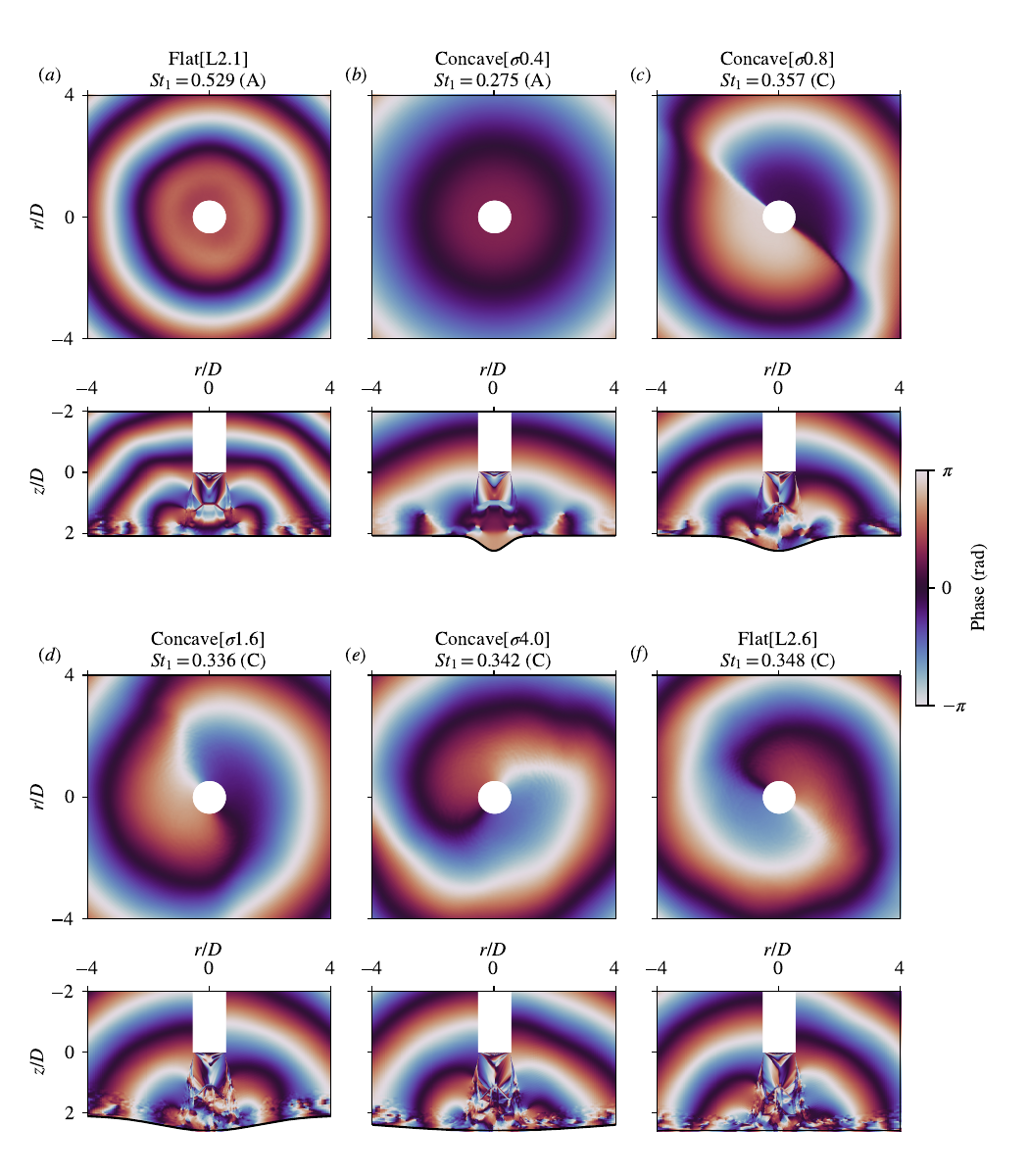}
    \caption{Phase fields of the density fluctuation field at each configuration's primary Strouhal number $St_1$ (table~\ref{tab:tones}): (a) Flat[L2.1], (b) Concave[$\sigma$0.4], (c) Concave[$\sigma$0.8], (d) Concave[$\sigma$1.6], (e) Concave[$\sigma$4.0] and (f) Flat[L2.6]. For each configuration the upper sub-panel is the nozzle-exit plane at $z=0$ and the lower sub-panel the meridional $r$-$z$ plane, with radial coordinate $r/D$ and axial coordinate $z/D$ increasing downward from the nozzle exit. The phase plotted is the argument of the temporal Fourier coefficient of the density; the solid black line in the meridional sub-panels marks the impingement surface.}
    \label{fig:phase}
\end{figure}

To determine the oscillation mode associated with the dominant tone of each configuration, a Fourier transform in time is applied to the density fluctuations on the nozzle-exit plane and on a meridional $r$-$z$ plane. The complex coefficient at the primary Strouhal number $St_1$ is retained, and its phase distinguishes the azimuthal mode orders \citep{powell1992observations, edgington2021generation}. The resulting phase fields are presented in figure~\ref{fig:phase}.

On the nozzle-exit plane, shown in the upper sub-panel for each configuration in figure~\ref{fig:phase}, Flat[L2.1] and Concave[$\sigma$0.4] are characterised by concentric or nearly azimuthally invariant phase, corresponding to an axisymmetric, toroidal mode of order $m=0$, hereafter denoted mode~A. In Flat[L2.1], multiple concentric phase rings indicate a finite radial acoustic wavelength on the exit plane, consistent with an outward-propagating acoustic wave at the screech frequency. In Concave[$\sigma$0.4], the phase varies only weakly across the disk, indicating a longer acoustic wavelength on the exit plane at the lower Strouhal number, $St=0.275$. The remaining four configurations, Concave[$\sigma$0.8], Concave[$\sigma$1.6], Concave[$\sigma$4.0], and Flat[L2.6], instead exhibit a single-arm logarithmic spiral pattern, characteristic of a helical mode of azimuthal order $m=\pm 1$, denoted mode~C. This separation of the six cases into mode~A and mode~C is summarised in table~\ref{tab:tones}.

Read alongside the Mach-disk dynamics of \S\ref{sec:machstem}, this classification reveals a clear correlation across the six cases: the two mode-A configurations, Flat[L2.1] and Concave[$\sigma$0.4], are precisely those that sustain a strong, periodic, axisymmetric Mach-disk pulsation (figure~\ref{fig:MachStem}), whereas the four mode-C configurations show only weak, broadband Mach-disk motion. This correlation is consistent with the established mode-dependence of shock-cell motion in screeching jets, in which axisymmetric modes drive an axial shock oscillation while helical modes drive an azimuthal one \citep{panda1998shock,edgington2019aeroacoustic}. The mechanism is a matter of symmetry. A toroidal ($m=0$) shear-layer disturbance reaches the Mach disk in phase around the azimuth and drives a net streamwise breathing motion of the shock. A helical ($m=\pm1$) disturbance reaches it out of phase, and therefore produces no coherent axial displacement (\S\ref{sec:machstem}). The resulting streamwise shock oscillation, locked to the screech cycle and most pronounced under the axisymmetric mode, is well documented for underexpanded jets \citep{panda1998shock,edgington2014coherent,li2023source}.

Among the concave cases, the transition from mode~A to mode~C coincides with the broadening of the indentation relative to the jet column: Concave[$\sigma$0.4], with $r_{99}/D \approx 1.2$, is axisymmetric, whereas the broader indentations, with $r_{99}/D \gtrsim 2.4$, are helical. The flat-wall pair shows, however, that standoff distance alone can also change the selected mode, since Flat[L2.1] is axisymmetric while Flat[L2.6] is helical. The mode selection therefore cannot be attributed to indentation geometry alone, and the present six cases are insufficient to isolate the controlling parameter. Azimuthal mode selection in impinging underexpanded jets is also sensitive to boundary conditions beyond the standoff distance, the nozzle external geometry being one documented control \citep{weightman2019nozzle}.

\begin{figure}
    \centering
    \includegraphics[width=\textwidth]{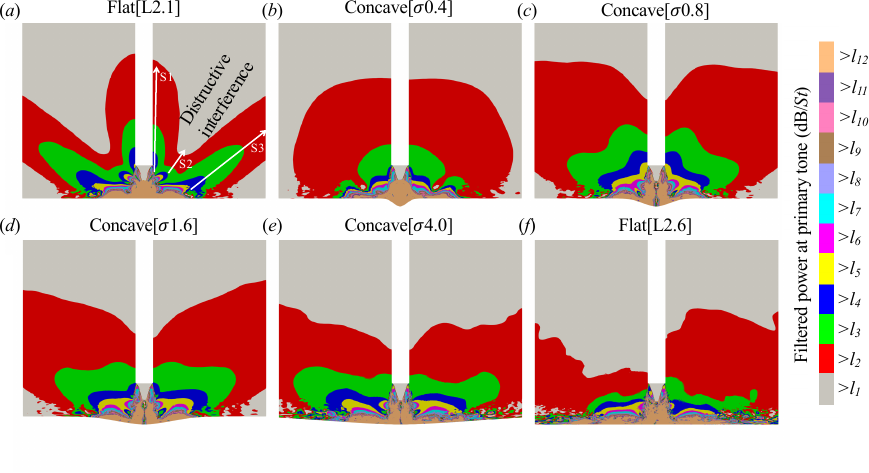}
    \caption{Amplitude fields filtered at the dominant screech-tone frequency, shown as twelve equal acoustic-level bands from $l_{1}$ to $l_{12}$ in dB$/$St; the green region, for example, denotes values within $[l_{3},\,l_{4}]$. The $l_{1}$ to $l_{12}$ range is set separately for each case to bring out the near-field directivity pattern.}
    \label{fig:directivity}
\end{figure}

The amplitude of the band-pass-filtered density-fluctuation field at the dominant Strouhal number is shown in figure~\ref{fig:directivity}, contoured in twelve equal acoustic-level intervals to highlight the directional structure of the radiated tone. Three coherent acoustic source regions are identified along the jet column for Flat[L2.1] (figure~\ref{fig:directivity}a), labelled S1, S2 and S3 in order of increasing radial offset. S1 is the upstream-pointing radiation emerging from the Mach disk along the jet axis. S2 and S3 are associated with the third and second shock cells respectively, the latter producing the radially-pointing lobe. The angular gaps between the lobes correspond to the loci where the contributions from S1-S3 interfere destructively. This lobed multipole pattern is characteristic of a toroidal screech mode formed by the coherent superposition of acoustic sources at successive shock-cell locations \citep{powell1953mechanism,harperbourne1973noise}, consistent with the multiple screech source locations identified along the shock cells by \citet{davies1962tones} and \citet{umeda2001sound}.

The radiation pattern of Concave[$\sigma$0.4] (figure~\ref{fig:directivity}b) lacks the angular gaps that characterise the Flat[L2.1] case and instead exhibits a single broad lobe enveloping the entire upstream half-space. This pattern is consistent with the dominance of a single coherent acoustic source, the unsteady Mach disk, whose radiated wavefronts are nearly spherical at the corresponding Strouhal number, in agreement with the meridional-plane phase distribution of figure~\ref{fig:phase}. The Concave[$\sigma$0.8] configuration (figure~\ref{fig:directivity}c) shows an intermediate behaviour, with a lobed primary-cell pattern superimposed on a broader rounded envelope. The three widest-spread configurations (figure~\ref{fig:directivity}d-f) display lobed patterns that are spatially confined to the immediate vicinity of the jet column and exhibit a substantially reduced upstream penetration of the tonal acoustic energy. The reduced amplitude and spatial extent of the tonal radiation in these cases is consistent with the order-of-magnitude difference in primary-tone SPL between the most and least concave configurations reported in figure~\ref{fig:SPL}.

\subsection{Powell-Tam source-transfer budget} \label{sec:why}


Sections~\ref{sec:nearfield} and~\ref{sec:modes} showed that the primary-tone sound-pressure level is highly sensitive to the wall curvature (figure~\ref{fig:SPL} and the $\mathrm{SPL}_1$ column of table~\ref{tab:tones}). This sensitivity is not obvious, because the upstream development of the jet is largely unaffected by the downstream wall (\S\ref{sec:flowfield}); the wall must therefore amplify the tone by acting on the feedback loop through the impingement region, rather than by modifying the incoming shear layer. The goal of this subsection is to identify which leg of the loop, the downstream source at the Mach disk or the upstream return of the acoustic wave to the nozzle, carries this curvature effect. In Powell's description \citep{powell1953mechanism,edgington2019aeroacoustic}, a self-sustained tone requires that a disturbance return to its origin no weaker than it began after one circuit of the loop. The round-trip gain must therefore satisfy
\begin{equation}\label{eqn:loop_gain}
q_d\,\eta_g\,\eta_u\,\eta_r \;\ge\; 1 ,
\end{equation}
where the four terms act in sequence around the loop. The downstream gain $q_d$ is the amplification of the Kelvin-Helmholtz wavepacket along the shear layer. The generation efficiency $\eta_g$ measures how effectively that wavepacket radiates a tonal acoustic source at the Mach disk. The transmission efficiency $\eta_u$ measures how much of the resulting upstream-propagating wave reaches the nozzle. The receptivity $\eta_r$ measures how effectively the returning wave excites a new instability wave near the lip. In the present LES, the lip has a finite thickness of $0.05D$, so the receptivity zone is the radial band $r/D\in[0.5,0.55]$ on the $z=0$ plane.

The jet-exit conditions are identical across the six cases, so the downstream gain $q_d$ is taken to be approximately the same in all of them; this assumption is revisited in the limitations discussion toward the end of this section. The curvature-induced amplification is then attributed to the remaining terms, which fall into two pathways whose relative importance in non-ideally expanded impinging jets remains unresolved. The first is the generation efficiency $\eta_g$ at the Mach disk, where the shock-vortex interaction radiates the tonal wave. The second is the upstream leg, $\eta_u\eta_r$, along which the returning disturbance may follow either of two physically distinct routes. It may propagate as a GJM trapped within the plume, with dispersion set by the shear-layer profile \citep{tam1990theoretical,bogey2017feedback}. Alternatively, it may propagate as an acoustic wave through the ambient outside the plume. A concave wall can redirect either of the two, and we refer to that redirection as `geometric focusing': the curved surface reflects the returning sound in the manner of a concave mirror, so that the indentation spread determines how much of the reflected energy reaches the lip receptivity zone. Which route carries the redirected wave depends on how far the indentation extends beyond the jet column, since a wall that curves only beneath the column returns the wave through the plume rather than through the ambient. The analysis below obtains a first-order partition of the curvature-induced amplification of the primary tone ($\Delta\mathrm{SPL}_{\rm lip}$) between the Mach-disk source and the upstream leg, and measures the upstream leg separately inside the plume and in the ambient.

The band-pass-filtered acoustic pressure at the primary tone $St_1$ is written as $p'_{\rm lip}=T_u\,p'_{\rm src}$, where $T_u\equiv\eta_u\,\eta_r$ is a lumped upstream-transfer function and $p'_{\rm src}$ a source-region pressure amplitude. Let $\Delta\mathrm{SPL}_{\rm lip}$ denote the primary-tone level of each configuration, in dB, relative to the flat-wall reference Flat[L2.6]. The curvature-induced amplification can then be partitioned additively in dB:
\begin{equation}\label{eqn:decomp}
\Delta\mathrm{SPL}_{\rm lip} \;=\; \Delta\mathrm{SPL}_{\rm src}^{\rm PT} \;+\; \Delta T_u.
\end{equation}
The source term $\Delta\mathrm{SPL}_{\rm src}^{\rm PT}$ is evaluated from the measured Mach-disk kinematics as
\begin{equation} \label{eqn:src_pt_db}
\Delta\mathrm{SPL}_{\rm src}^{\rm PT} = 20\log_{10}\!\left[\frac{z_{\rm ms}^{\rm rms}\,St_1}{\left(z_{\rm ms}^{\rm rms}\,St_1\right)_{\rm ref}}\right],
\end{equation}
where $z_{\rm ms}^{\rm rms}$ is the total root-mean-square axial Mach-disk displacement and the subscript `${\rm ref}$' denotes the Flat[L2.6] reference. This expression follows from the Powell-Tam kinematic scaling for shock-vortex tonal radiation \citep{powell1953mechanism,tam1990theoretical}: for a shock front of rms axial displacement $z_{\rm ms}^{\rm rms}$ oscillating at the primary tone frequency $f_1=St_1\,u_j/D$, the radiated tonal pressure follows
\begin{equation} \label{eqn:powell_tam}
p'_{\rm src} \;\propto\; \rho_\infty\,a_\infty\,z_{\rm ms}^{\rm rms}\,D\,(2\pi St_1\,u_j/D),
\end{equation}
so the source amplitude scales as $z_{\rm ms}^{\rm rms}\,St_1$.
This kinematic prediction depends only on the measured Mach-disk rms reported in \S\ref{sec:machstem} and on the measured primary-tone Strouhal in table~\ref{tab:tones} (both are direct LES outputs). The transfer term $\Delta T_u=\Delta\mathrm{SPL}_{\rm lip}-\Delta\mathrm{SPL}_{\rm src}^{\rm PT}$ is then a dB residual; by construction it absorbs any error in the source estimate as well as any genuine variation in $\eta_u\,\eta_r$, and the residual interpretation is therefore not by itself evidence of geometric focusing. To address this limitation, the upstream-propagating wave is measured directly in the LES pressure field on the $r$-$z$ plane. A concave wall can redirect the returning wave through the plume as well as through the ambient, and for the narrow indentations most of the curvature lies beneath the jet column: $78\%$ of the Concave[$\sigma$0.4] indentation, weighted by area, falls inside $r/D=0.7$. The measurement is therefore made in two radial bands, the jet column $r/D\in[0,0.5]$ and the ambient $r/D\in[0.7,3.0]$, giving $\Delta\mathrm{SPL}_{\rm up}^{\rm int}$ and $\Delta\mathrm{SPL}_{\rm up}^{\rm ext}$. The shear layer between them is excluded from both bands so that the two channels remain separate. For each case, the pressure $p(z, r, t)$ on the meridional plane is restricted to the axial window $z/D\in[0, 2.0]$, and for each $r$-slice a $(z, t)$ Fourier decomposition is performed after Hann$\times$Hann windowing. The spectral power is integrated over a mask that selects the upstream half-plane $k_z<0$ and a $\pm 0.02$ Strouhal band around the case's $St_1$, giving the per-$r$ upstream tonal energy. Summing over the $r$-slices on both sides of the jet axis gives the band total.

\begin{figure}
    \centering
    \includegraphics[width=\textwidth]{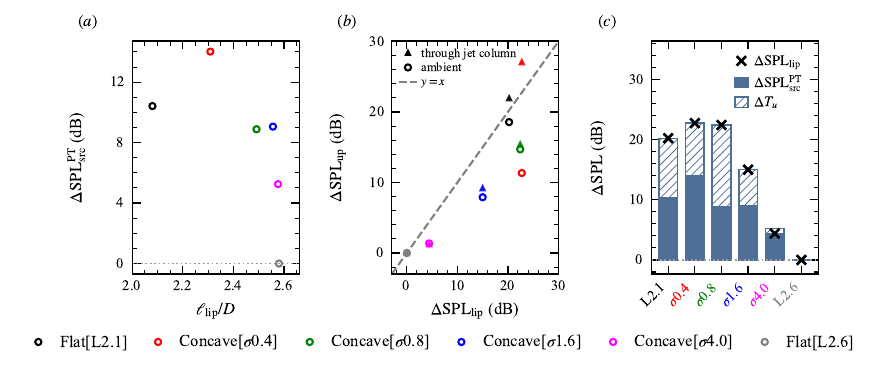}
    \caption{Decomposition of the spread-dependent primary-tone amplification. (a) Powell-Tam source-strength amplification $\Delta\mathrm{SPL}_{\rm src}^{\rm PT}$ equation~(\ref{eqn:src_pt_db}) versus the lip-line feedback length $\ell_{\rm lip}/D$. (b) Upstream-wave amplification in the jet column, $r/D\in[0,0.5]$ (filled triangles), and in the ambient, $r/D\in[0.7,3.0]$ (open circles), against the measured tone amplification $\Delta\mathrm{SPL}_{\rm lip}$, with the $y=x$ line. (c) Per-case attribution: stacked $\Delta\mathrm{SPL}_{\rm src}^{\rm PT}$ (blue) and $\Delta T_u$ (orange) bars, with crosses marking the measured $\Delta\mathrm{SPL}_{\rm lip}$.
    }
    \label{fig:why_curvature}
\end{figure}

Figure~\ref{fig:why_curvature} reports the decomposition of equation~(\ref{eqn:decomp}), referenced to Flat[L2.6]. Panel~(a) plots the source term against the lip-line feedback length $\ell_{\rm lip}/D$. It is largest for the narrowest indentation, $+14.0$~dB at Concave[$\sigma$0.4], falls to $+5.3$~dB at Concave[$\sigma$4.0], and is indistinguishable at the two intermediate spreads, $+8.9$ and $+9.1$~dB. Flat[L2.1] carries $+10.4$~dB despite being flat, its shorter standoff raising the Mach-disk rms, so the source term embeds an impingement-distance dependence as well as a spread dependence. Panel~(b) plots both upstream channels against the measured tone amplification. Both grow with the tone, and the column channel is the larger in every configuration, reaching $+27.1$~dB against $+11.3$~dB in the ambient at Concave[$\sigma$0.4]. The two energies together differ from the column value by at most $0.2$~dB, so the upstream tonal field is carried predominantly within the plume. The column amplitude follows the $y=x$ line to within $7.0$~dB and the ambient to within $11.5$~dB. The abscissa is $\Delta\mathrm{SPL}_{\rm lip}$ rather than the residual $\Delta T_u$ because $\Delta T_u$ is a change in a transfer ratio while the measurement is a change in an absolute level, and the two coincide only for a source common to all cases. Isolating $T_u$ would require the ratio of the upstream amplitude at the lip to that at the source, and an axial window short enough to localise the lip cannot resolve the sign of $k_z$.

Panel~(c) stacks the source and transfer contributions against the measured $\Delta\mathrm{SPL}_{\rm lip}$. For the two most strongly amplified configurations the split is $62\%/38\%$ (source/transfer) at Concave[$\sigma$0.4] and $40\%/60\%$ at Concave[$\sigma$0.8]. At the weakly indented Concave[$\sigma$4.0] the Powell-Tam source term alone slightly exceeds the measured $\Delta\mathrm{SPL}_{\rm lip}$, leaving a small negative transfer residual, so the additive split is informative only where the amplification is large. Three inferences follow.

\begin{list}{(\roman{enumi})}{%
  \usecounter{enumi}%
  \setlength{\leftmargin}{2.2em}\setlength{\labelwidth}{1.8em}%
  \setlength{\labelsep}{0.4em}\setlength{\itemindent}{0pt}%
  \setlength{\listparindent}{0pt}\setlength{\topsep}{4pt}%
  \setlength{\itemsep}{3pt}\setlength{\parsep}{0pt}%
  \def\makelabel#1{#1\hfil}}
\item The amplification of the primary tone is shared between the Mach-disk source and the upstream leg of the feedback loop in proportions that vary with the indentation spread, the source carrying $62\%$ of it at Concave[$\sigma$0.4] against $40\%$ at Concave[$\sigma$0.8].
\item The upstream-propagating wave strengthens together with the tone in the jet column and in the ambient alike, so the transfer term reflects a real increase in the returning wave rather than an artefact of its construction as a residual.
\item The upstream field is column-dominated in every configuration, the column exceeding the ambient by $14$ to $30$~dB. It is also the column that gains most as the indentation narrows, by $15.8$~dB more than the ambient at Concave[$\sigma$0.4] against $1.3$~dB or less at the wider ones, so the wall acts on the returning wave through the plume while its curvature remains beneath the jet column, and through the ambient once it extends beyond.
\end{list}

The second and third of these are corroborated in \S\ref{sec:waves_pod} and \S\ref{sec:dispersion}, where the same upstream field is characterised directly from its radial structure and its dispersion. That the external field participates is consistent with direct observation of an upstream-propagating wave outside the plume in impinging jets \citep{edgington2019aeroacoustic}, and with the demonstration that such a wave can close the loop after reflecting from an upstream surface \citep{weightman2019nozzle}, while the dominance of the column channel is consistent with an upstream wave confined to the plume \citep{tam1990theoretical,bogey2017feedback}.

\noindent\textit{Limitations.}\;The decomposition of equation~(\ref{eqn:decomp}) comes with limitations. The Powell-Tam scaling represents the source amplitude through the Mach-disk rms displacement and the tone frequency alone. Curvature-induced variations in shock strength and angle, vortex amplitude and source directivity are not represented, and none is separately constrained here, so the source term carries an uncertainty the budget cannot quantify. The displacement entering equation~(\ref{eqn:src_pt_db}) is the total rms of the tracked axial position, whereas equation~(\ref{eqn:powell_tam}) is a scaling for tonal radiation. For Flat[L2.1] and Concave[$\sigma$0.4] the distinction is immaterial, since restricting the record to a band of half-width $\Delta St = 0.02$ about the primary tone retains $84\%$ and $93\%$ of the rms. For the four helical configurations it retains only $5$-$10\%$ (\S\ref{sec:machstem}). There the source term measures broadband shock excursion, and the axial motion is in any case not the tonal source, since a helical oscillation displaces the shock azimuthally rather than axially (\S\ref{sec:modes}). The source and transfer shares therefore rest on firmer ground for the axisymmetric configurations. The feedback loop is bidirectional, the source amplitude depending on a gain that itself depends on $T_u$, so equation~(\ref{eqn:decomp}) is best read as an instantaneous balance of the saturated limit cycle rather than as cause and effect. The upstream measurement constrains the sign and the ordering of the transfer term but not its magnitude, and the column and ambient contributions to $\eta_u\,\eta_r$ cannot be apportioned from it, since $\Delta T_u$ is a residual rather than a measured ratio. The shear-layer gain $q_d$ is assumed common to all cases, supported by the near-invariant convection velocity of \S\ref{sec:Uc}. The Flat[L2.6] reference differs in mean standoff as well as in shape, so the spread attribution retains a standoff contribution. A direct comparison of the upstream-wave amplitude across the six configurations on a common scale is presented in \S\ref{sec:waves_pod}.

\section{Loading on the impinging surface} \label{sec:loading}
The unsteady Mach disk (\S\ref{sec:machstem}) and the toroidal/helical jet-column oscillations (\S\ref{sec:modes}) translate into a time-varying force and moment distribution on the impinging surface. The spectral content of this loading is therefore expected to be inherited from the oscillation modes. The modal organisation of the surface loads has received little attention; prior studies characterised mainly the mean wall pressure and overall load \citep{Donaldson_Snedeker_1971, lamont1980impingement, prasad1994impingement}. Unsteady wall-pressure fluctuations on a curved impingement plate were measured by \citet{ho1977surface}, but their modal organisation was not examined. This section addresses three questions: (i) the modal composition of the unsteady surface forces and moments, (ii) the correspondence between these loading modes and the jet-column screech modes, and (iii) the dependence of the loading on wall curvature. The wall-shear topology and the integrated forces are examined in \S\ref{sec:forces}; the moments about the surface centroid, which measure the spatial asymmetry of the wall loading, are addressed in \S\ref{sec:moments}.

\subsection{Forces and stresses} \label{sec:forces}

\begin{figure}
    \centering
    \includegraphics[width=\textwidth]{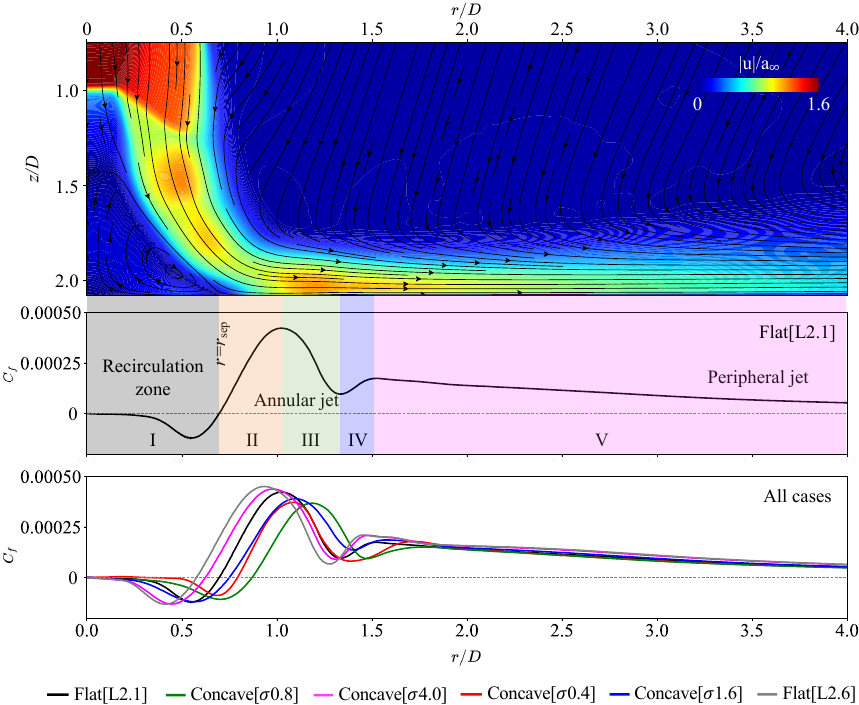}
    \caption{Wall-flow topology and skin friction on the impinging surface.
    Top: mean velocity magnitude $\overline{|\mathbf{u}|}/a_{\infty}$ and projected streamlines in the meridional $r$-$z$ plane within $0 \le r/D \le 4$ for the Flat[L2.1] configuration.
    Middle: radial distribution of the skin friction coefficient $C_f$ for Flat[L2.1], with annotations identifying the recirculation zone (I), the annular jet (II-III), the relaxation region (IV), and the peripheral wall jet (V). The reattachment radius $r=r_{\text{sep}}$ marks the zero-crossing of $C_f$.
    Bottom: $C_f(r)$ for all six configurations, illustrating the outward shift of the reattachment radius and the broadening of the annular-jet footprint with decreasing indentation spread $\sigma$.
    }
    \label{fig:skin_friction}
\end{figure}

The mean near-wall flow, shown in the top panel of figure~\ref{fig:skin_friction}, consists of three regions: an inner stagnation/recirculation bubble, a radially accelerated annular wall jet that turns the impinging shear layer through nearly $90^{\circ}$, and a thinning peripheral jet. The skin-friction coefficient, $C_f = \tau_w / (\tfrac{1}{2}\rho_j u_j^2)$, follows this organisation and divides into five radial zones I-V (middle panel). Zone I is the reversed-flow recirculation bubble, where $C_f<0$, and ends at the reattachment radius $r_{\text{sep}}$. Zones II-III correspond to the annular jet, where $C_f$ peaks at $\approx 4.5\times 10^{-4}$ over $0.7 \lesssim r/D \lesssim 1.3$. Zones IV-V mark the relaxation of the annular jet and the decay of the peripheral wall jet.

The comparison across the six cases in the bottom panel of figure~\ref{fig:skin_friction} shows that this five-zone organisation is retained, but the radial extent of each zone changes with wall geometry. The reattachment radius $r_{\text{sep}}$ does not vary in a single direction with the indentation spread: from $r/D \approx 0.70$ for the baseline flat wall Flat[L2.1] it grows through Concave[$\sigma$0.4] to a maximum of $\approx 0.87$ for Concave[$\sigma$0.8], and then contracts steadily across the broader indentations to $\approx 0.58$ for Flat[L2.6]. The recirculation zone is thus widest at an intermediate spread, mirroring the enlarged recirculation pocket behind the Mach disk identified in \S\ref{sec:flowfield}. The peak in $C_f$ tracks this variation inversely: it is smallest for the two cases with the largest recirculation bubble, Concave[$\sigma$0.8] and Concave[$\sigma$0.4] ($\approx 3.7\times10^{-4}$), and largest for the flat and widest-spread walls, Flat[L2.6] and Concave[$\sigma$4.0] ($\approx 4.5\times10^{-4}$), whose bubbles are smallest. The peak also moves radially outward as the bubble widens, from $r/D \approx 0.9$ for Flat[L2.6] to $r/D \approx 1.1$-$1.2$ for the two narrowest indentations. A larger recirculation bubble therefore spreads the impinging momentum over a broader annulus and lowers the peak wall shear, consistent with radial deflection of the wall jet by the concave indentation.

\begin{figure}
    \centering
    \includegraphics[width=\textwidth]{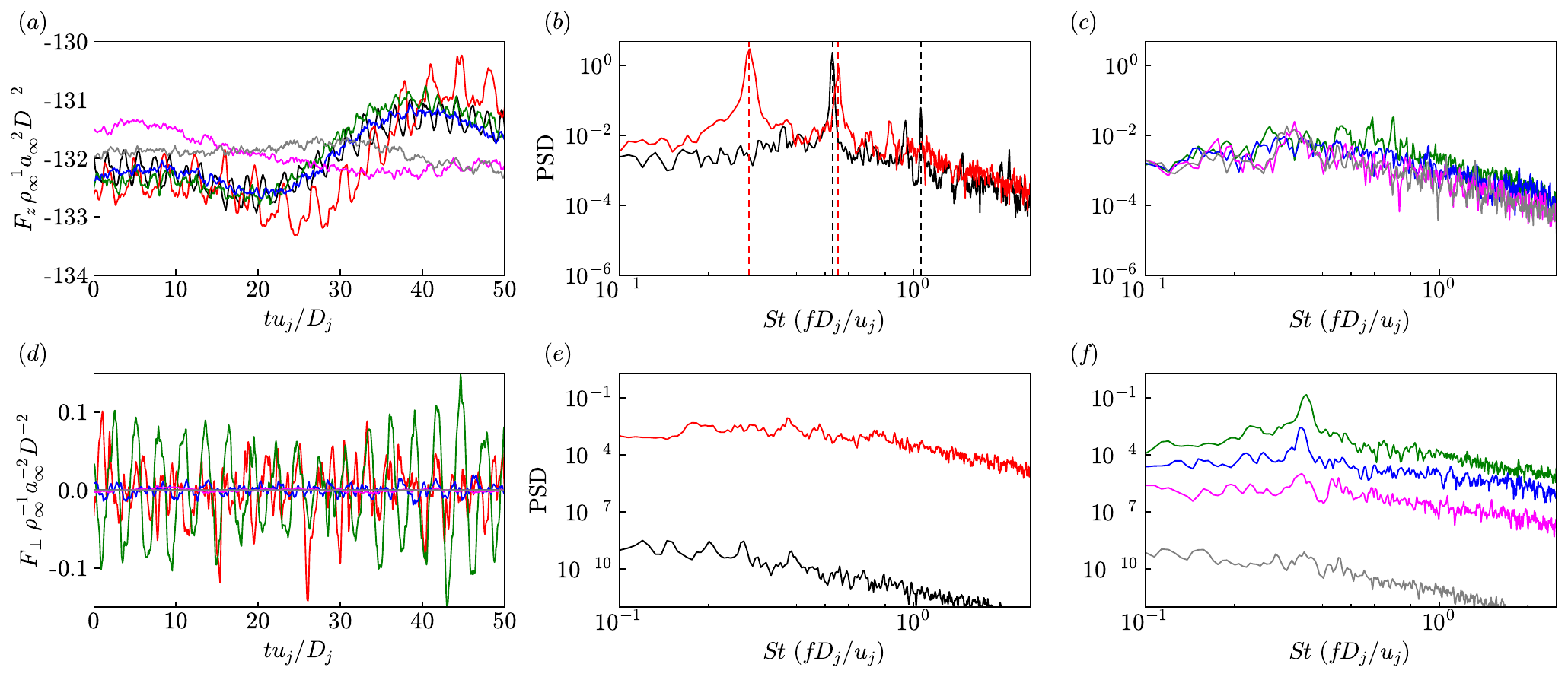}
    \caption{Axial and transverse forces on the impinging surface.
    (a) Time history of the axial force $F_z/(\rho_{\infty}a_{\infty}^2 D^2)$.
    (b,c) Power spectral density (PSD) of $F_z$ for (b) Flat[L2.1] (black) and Concave[$\sigma$0.4] (red), and (c) the remaining four configurations. Vertical dashed lines in (b) mark the dominant screech tones reported in table~\ref{tab:tones}.
    (d) Time history of the transverse (in-plane) force $F_{\perp}/(\rho_{\infty}a_{\infty}^2 D^2)$.
    (e,f) PSD of $F_{\perp}$ grouped as in (b,c). The tone at $St=0.357$ for Concave[$\sigma$0.8] (green) in panel~(f) coincides with the dominant tonal Strouhal of that configuration.
    Colour coding follows figure~\ref{fig:fft_probes}.
    }
    \label{fig:forces}
\end{figure}

The azimuthal symmetry of the wall pressure determines which component of the net loading carries the screech tone, independently of the detailed radial pressure distribution. Decomposing the fluctuating wall pressure on the impingement surface into azimuthal Fourier modes about the jet axis gives
\begin{equation} \label{eqn:azim_projection}
    p'(r,\theta,t) = \sum_{m} \hat{p}_m(r,t)\,e^{\mathrm{i}m\theta},
    \qquad
    F_z'(t) = -\!\int_0^{2\pi}\!\!\!\int_0^{R}\! p'\,r\,\mathrm{d}r\,\mathrm{d}\theta
    = -2\pi\!\int_0^{R}\! \hat{p}_0(r,t)\,r\,\mathrm{d}r,
\end{equation}
so that the net axial force follows from the axisymmetric mode alone. Here $r$ and $\theta$ are the surface-radial and azimuthal coordinates, $R$ is the outer radius of the sampled surface, and the minus sign follows from the pressure acting along the inward surface normal $\mathbf{n}$. The wall curvature leaves no metric factor in the axial projection: for a surface $z=h(r)$ the element of axial projected area is $(\mathbf{n}\!\cdot\!\hat{\mathbf{z}})\,\mathrm{d}S = r\,\mathrm{d}r\,\mathrm{d}\theta$, independently of $h$. The second equality follows from the azimuthal orthogonality relation, $\int_0^{2\pi} e^{\mathrm{i}(m-m')\theta}\,\mathrm{d}\theta = 2\pi\delta_{mm'}$, which eliminates all $m\neq 0$ contributions to $F_z'$. By the same orthogonality, the transverse (in-plane) force $F_{\perp}$ and the in-plane bending moment $(M_{\perp,1},M_{\perp,2})$ are set solely by the first helical modes, $m=\pm 1$. The net axial loading therefore responds only to the axisymmetric pressure mode, while the transverse force and bending moment respond only to the helical mode. This is the surface-loading counterpart of the multipole classification of \citet{powell1953mechanism}, in which the $m=0$ and $m=\pm 1$ jet-column oscillations radiate as axial and lateral sources, respectively. A steadily rotating $m=1$ pattern, $\hat{p}_1\propto e^{-\mathrm{i}\omega t}$, drives a bending-moment vector $(M_{\perp,1},M_{\perp,2})$ that precesses about the jet axis at the tone frequency. The screech mode of each configuration (table~\ref{tab:tones}) thus selects, through equation~(\ref{eqn:azim_projection}), whether the tone appears in the axial force or in the transverse force and moment.

\textit{Axial forces.}\; The integrated axial and transverse forces on the impinging surface, obtained by area-integration of the wall pressure and viscous stress, are presented in figure~\ref{fig:forces}. All configurations fluctuate about a comparable mean, $\overline{F}_z/(\rho_{\infty}a_{\infty}^2 D^2) \approx -132$ (figure~\ref{fig:forces}a). This value is set almost entirely by the ambient pressure acting over the sampled surface, whose axial projected area is $181D^2$; removing that offset leaves a net axial load of $\approx -2.5$, which is the thrust reaction borne by the surface and agrees to within $7\%$ with the ideally expanded jet thrust. Measured against this net load the unsteady component is not small: the root-mean-square of $F_z$ ranges from $0.23$ for Flat[L2.6] to $1.08$ for Concave[$\sigma$0.4], that is from $9\%$ to $43\%$ of the mean net load, and the peak-to-peak excursion exceeds the mean net load for the four most strongly tonal configurations. The modal composition of this unsteady load follows the screech mode directly. The axial-force spectra (figure~\ref{fig:forces}b,c) separate the six cases into the two groups of the mode classification of table~\ref{tab:tones}: the two axisymmetric (mode-A) cases carry discrete peaks at their screech tone and its harmonic, whereas the four helical (mode-C) cases give a broadband $F_z$ with no emerging tone. This is the action of the projection of equation~(\ref{eqn:azim_projection}): only the $m=0$ pressure mode contributes to the net axial force, so $F_z$ registers the tone when the jet-column oscillation is axisymmetric and filters it out when the oscillation is helical.

\textit{Transverse forces.}\; The transverse (in-plane) force $F_{\perp}$ (figure~\ref{fig:forces}d) is orders of magnitude smaller than $F_z$, so the net surface loading is overwhelmingly axial. It nonetheless reflects the non-axisymmetric jet-column oscillations, and its magnitude depends strongly on the wall curvature. On a planar wall the pressure acts purely along the axis and exerts no net in-plane force, so $F_{\perp}$ is carried entirely by the weak wall shear; on a concave wall the tilted surface normal projects the wall pressure into the surface plane, and the pressure contribution then dominates. This geometric switch is evident in the spectra of figure~\ref{fig:forces}(e,f): the transverse load on the two flat walls is negligible, and introducing even the weakest indentation, Concave[$\sigma$4.0], raises the transverse fluctuation power by three to four orders of magnitude, rising to about six orders at the intermediate spreads. By the same orthogonality that isolates $F_z$, the transverse force responds only to the $m=\pm 1$ pressure mode, and it accordingly carries a sharp tone for the helical cases whose screech resides in that mode, the clearest being Concave[$\sigma$0.8] at its dominant Strouhal number. The axial and transverse forces thus verify the selection rule of equation~(\ref{eqn:azim_projection}): the transverse tone is the wall-loading counterpart of the lateral radiation of the $m=\pm 1$ jet-column oscillation.

\subsection{Moments on the impinging surface} \label{sec:moments}

\begin{figure}
    \centering
    \includegraphics[width=\textwidth]{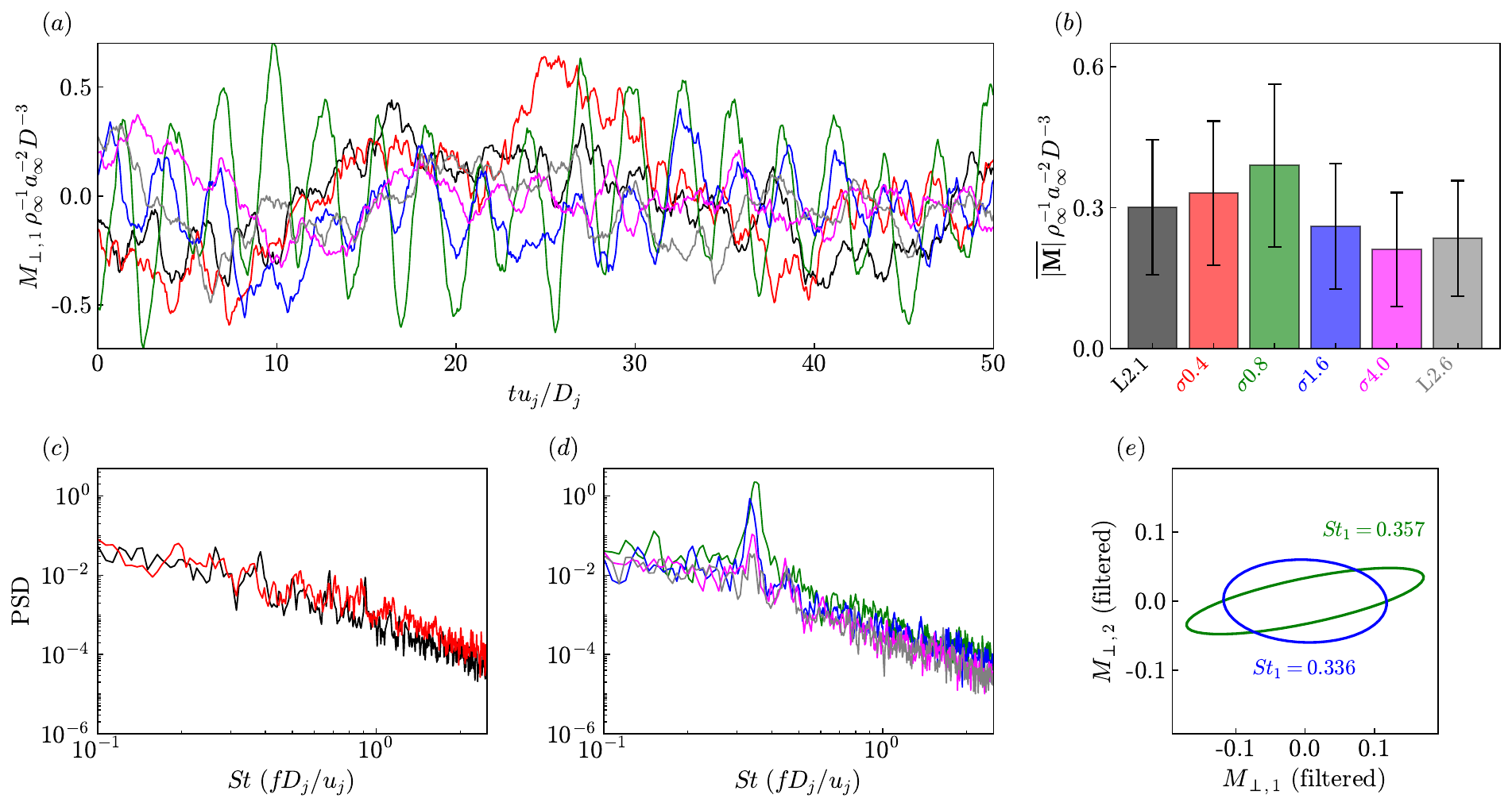}
    \caption{Moments about the centre of the impinging surface.
    (a) Time history of one in-plane bending-moment component, $M_{\perp,1}/(\rho_{\infty}a_{\infty}^2 D^3)$, for all six configurations.
    (b) Time-averaged moment magnitude $\overline{|\mathbf{M}|}/(\rho_{\infty}a_{\infty}^2 D^3)$ for each configuration; error bars indicate the root-mean-square fluctuation.
    (c,d) PSD of $M_{\perp,1}$ for (c) Flat[L2.1] (black) and Concave[$\sigma$0.4] (red), and (d) the remaining four configurations. Sharp tones at $St=0.357$ and $St=0.336$ in (d) match the dominant tonal Strouhal numbers of Concave[$\sigma$0.8] and Concave[$\sigma$1.6] (table~\ref{tab:tones}).
    (e) Trajectory of the in-plane bending moment $(M_{\perp,1},M_{\perp,2})$ band-pass-filtered at the dominant tonal Strouhal number for Concave[$\sigma$0.8] (green) and Concave[$\sigma$1.6] (blue); the closed elliptical orbits indicate a moment vector precessing about the jet axis in synchrony with the helical jet-column mode.
    Colour coding follows figure~\ref{fig:fft_probes}.
    }
    \label{fig:moments}
\end{figure}

The moments of the wall loading measure its spatial asymmetry and govern the bending and fatigue loading of the impinging structure. The moment vector is computed as
\begin{equation} \label{eqn:moment}
    \mathbf{M}(t) = \int_{S} (\mathbf{r}-\mathbf{r}_c) \times \mathbf{s}(\mathbf{r},t)\,\mathrm{d}S,
\end{equation}
where $\mathbf{r}_c$ is the surface centroid, $\mathbf{s}=-p\,\mathbf{n} + \boldsymbol{\tau}\cdot\mathbf{n}$ is the total wall stress vector, $\mathbf{n}$ the inward surface normal, and the moment is non-dimensionalised by $\rho_{\infty}a_{\infty}^2 D^3$. Because the wall pressure acts along the near-axial normal, it exerts only in-plane bending moments and negligible torsion about the jet axis; the twist component is smaller than the bending components by three orders of magnitude and is not considered further. The two in-plane components $(M_{\perp,1},M_{\perp,2})$ form the first radial moment of the wall pressure, which by the orthogonality of equation~(\ref{eqn:azim_projection}) is set by its $m=\pm 1$ azimuthal part. The bending moment is thus the helical-mode counterpart of the transverse force, but weighted by the lever arm about the centroid, so it is the more sensitive and fatigue-relevant measure of the non-axisymmetric loading.

The time histories of the in-plane bending moment in figure~\ref{fig:moments}(a) are zero-mean and oscillate at amplitudes of order $M_{\perp,1}/(\rho_{\infty}a_{\infty}^2 D^3) \approx 0.5$ in every configuration, and the time-averaged moment magnitude (figure~\ref{fig:moments}b) is largest at the intermediate spread Concave[$\sigma$0.8], about $30\%$ above the Flat[L2.1] baseline. Their spectra in figure~\ref{fig:moments}(c,d) establish the modal correspondence directly. The two axisymmetric cases, Flat[L2.1] and Concave[$\sigma$0.4], give a broadband $M_{\perp,1}$ with no tonal peak, since their screech resides in the $m=0$ pressure mode that carries no net bending moment. The helical cases instead concentrate the moment fluctuation at their screech frequency, with sharp tones at $St=0.357$ and $0.336$ for Concave[$\sigma$0.8] and Concave[$\sigma$1.6] (table~\ref{tab:tones}), while the nearly flat helical cases Concave[$\sigma$4.0] and Flat[L2.6] leave only a weak tonal residue. The tonal bending moment thus appears for exactly the configurations whose axial force is tonally silent, and vice versa. This complementarity is the moment-level counterpart of the selection rule of equation~(\ref{eqn:azim_projection}): the axial force and the in-plane moment form the $m=0$ and $m=\pm 1$ channels of the wall loading, and between them carry the tone of every screech mode.

The kinematics of the tonal moment are shown by the band-pass-filtered trajectory of $(M_{\perp,1},M_{\perp,2})$ in figure~\ref{fig:moments}(e), which forms closed elliptical orbits for Concave[$\sigma$0.8] and Concave[$\sigma$1.6]. The moment vector precesses about the jet axis at the tone frequency, imposing a rotating bending load on the surface. The orbit geometry is set by the azimuthal composition of the helical mode, which may be written as a superposition of two counter-rotating helical waves of order $m=+1$ and $m=-1$ \citep{powell1992observations,edgington2019aeroacoustic}. A single rotational sense gives a moment vector of constant magnitude and hence a circular orbit, corresponding to a spinning mode. Two counter-rotating waves of equal amplitude confine the vector to a line, corresponding to a standing, flapping mode. Unequal amplitudes such as in Concave[$\sigma$0.8] and Concave[$\sigma$1.6] give an ellipse \citep{gojon2019antisymmetric}. The moderate eccentricity of the present orbits therefore indicates a helical mode in which the two rotational senses have unequal amplitudes. A similar weak preference for one rotational sense has been reported for free screeching jets, where either sense occurs with near-equal probability but one carries slightly stronger fluctuations \citep{edgington2022screeching}. The relative amplitudes of the two waves may also vary in time, giving the flapping-spin-flapping behaviour observed in reconstructed density and velocity fields of underexpanded screeching jets \citep{lee2026experimental}. The closed orbit measured here is the time-averaged projection of that motion onto the wall loading.

\section{Modelling the downstream- and upstream-propagating waves} \label{sec:waves}

\subsection{Convection velocity} \label{sec:Uc}

\begin{figure}
\centering
\includegraphics[width=\textwidth]{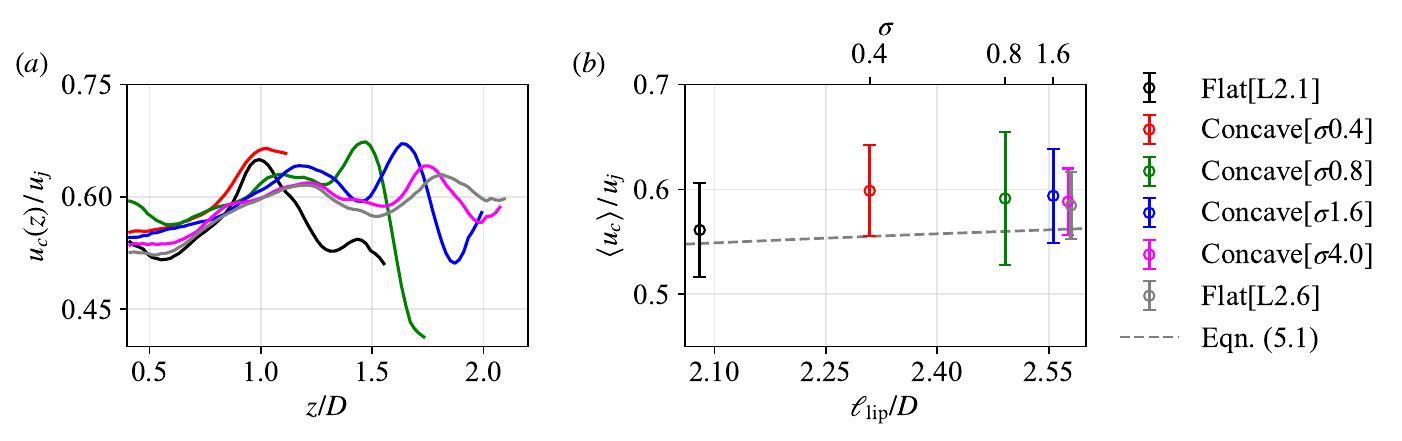}
\caption{Mean convection velocity of the downstream wavepacket along the maximum-turbulent-kinetic-energy shear-layer trajectories for the six configurations.
(a) Local $u_c(z)/u_j$ versus axial station $z/D$, averaged over the upper and lower trajectories.
(b) Trajectory-mean $\langle u_c \rangle/u_j$ for each configuration, obtained by integrating the averaged profile along $z$, plotted against the lip-line feedback length $\ell_{\rm lip}/D$ from equation~(\ref{eqn:ell_lip}) on the lower axis; the indentation spread $\sigma$ of the concave cases is marked on the aligned upper axis. The dashed line gives the empirical relation of equation~(\ref{eqn:gb_uc}).}
\label{fig:uc_xcorr}
\end{figure}

The mean convection velocity $\langle u_c \rangle$ of the downstream-propagating wavepacket is estimated directly from the LES on the meridional $r$-$z$ plane, using an axial cross-correlation of shear-layer velocity signals, following the approach used for round impinging jets by \citet{gojon2017flow}. For each configuration, the upper and lower shear layers are traced as the loci of maximum turbulent kinetic energy from the nozzle lip to the impingement region. The axial velocity along these trajectories is then band-pass filtered over the jet-based Strouhal band $St \in [0.1,\,1.5]$ to isolate the large-scale convecting wavepacket from finer-scale turbulence and the faster acoustic field. The local convection velocity is obtained from the peak-correlation lag between stations separated by $\Delta z = 0.25D$, retaining only estimates in the physical range $[0.40,0.70]u_j$. For the most concave configuration, Concave[$\sigma$0.4], no coherent convecting wavepacket can be tracked beyond $z/D \approx 1.1$, where the shear-layer ridge is overtaken by a near-standing feedback and shock-oscillation field. Its profile in figure~\ref{fig:uc_xcorr}(a) therefore terminates upstream of the other cases. These convection velocities should be interpreted as estimates, since the short timescales of the convecting structures and the limited usable streamwise extent both constrain the accuracy of the lag-based extraction.

\citet{gojon2017flow} showed that the mean convection velocity in a round impinging jet is well represented by
\begin{equation}\label{eqn:gb_uc}
\frac{\langle u_c \rangle}{u_e}
=
0.65\frac{u_j}{u_e}
-
\left(0.65\frac{u_j}{u_e} - 0.5\right)
\frac{1}{1 + L/D_j},
\end{equation}
which tends to $0.65u_j$ for $L\gg D_j$ and to $0.5u_e$ for $L\ll D_j$. For the present underexpanded sonic exit, with $M_e=1$, $u_j/u_e \approx 1.40$ for $M_j=1.56$ and $\gamma=1.4$, and $D_j/D \approx 1.10$, equation~(\ref{eqn:gb_uc}) gives $\langle u_c \rangle/u_j=0.549$ at $L/D=2.1$ and $0.563$ at $L/D=2.6$. The trajectory-based estimates are $\langle u_c \rangle/u_j = 0.561 \pm 0.045$ for Flat[L2.1] and $0.585 \pm 0.032$ for Flat[L2.6], both within about $0.02u_j$ of the empirical prediction. The four concave configurations cluster between $0.588$ and $0.599$, slightly above the flat-wall values and nearly independent of the indentation spread $\sigma$.

The convection velocity shows weak sensitivity to the wall geometry, remaining between $0.56$ and $0.60u_j$ for all six configurations. This weak variation is consistent with $\langle u_c \rangle$ being set primarily by the initial shear-layer development near the nozzle lip, where the geometry and jet Mach number are common to all cases and the flow has not yet responded strongly to the downstream wall. Thus, the wall curvature modifies the impingement region without substantially changing the upstream shear-layer wavepacket that forms the downstream leg of the feedback loop. This wavepacket also carries the downstream gain $q_d$ of the loop-gain balance of equation~(\ref{eqn:loop_gain}), so its near-invariance across the six cases supports the assumption of \S\ref{sec:why} that $q_d$ is common to all configurations, localising the curvature-induced amplification to the Mach-disk source and the upstream leg rather than the downstream convection. The per-case values of $\langle u_c \rangle$ reported in table~\ref{tab:st1_predictions} are therefore used directly in the modified Powell feedback equation of \S\ref{sec:feedback}. Given the uncertainty of the lag-based extraction, these values are used only to set the downstream convection time in the feedback-loop model.


\subsection{Wavenumber spectrum and spatial structure of the upstream-propagating wavepacket}
\label{sec:waves_pod}

\begin{figure}
    \centering
    \includegraphics[width=\textwidth]{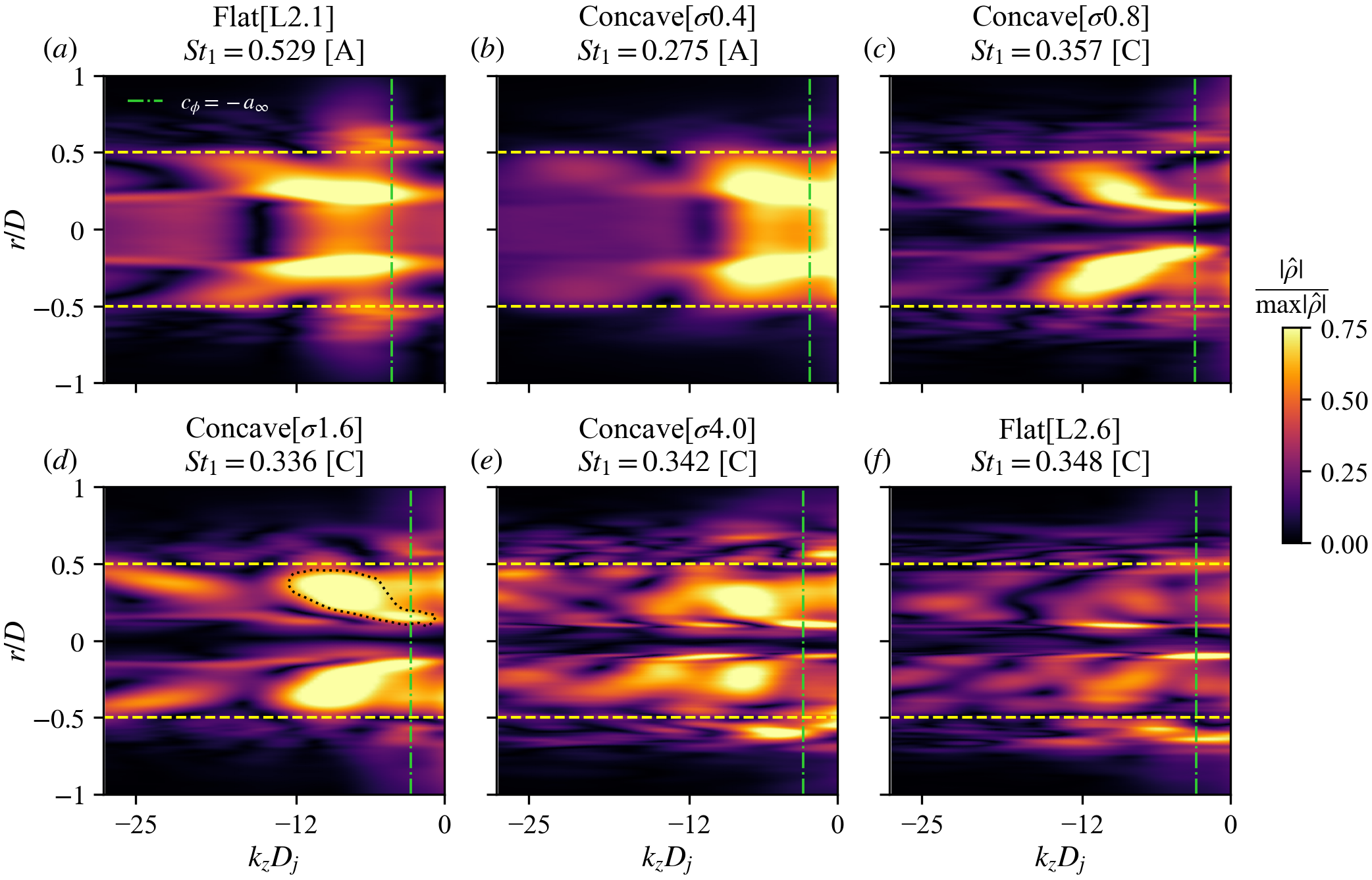}
    \caption{Wavenumber spectrum $|\hat\rho(k_z D_j, r/D)|$ at each case's primary tone $St_1$ on the meridional $r$-$z$ plane, restricted to the upstream-propagating half $k_z D_j < 0$ and normalised in each panel by its own maximum. The yellow dashed lines mark the lip line at $|r|/D = 0.5$, and the green dash-dot line marks $k_z = -k_a$, with $k_a = 2\pi f_1/a_\infty$, at which the upstream phase speed equals the ambient sound speed in magnitude ($c_\phi = -a_\infty$). The white dashed curve in panel~(d) outlines one of the two petals of the helical cases.}
    \label{fig:kx_zd}
\end{figure}

The upstream leg of the feedback loop may close either through a guided jet mode trapped within the plume \citep{tam1990theoretical,bogey2017feedback} or through an externally propagating acoustic wave radiating through the ambient. These two routes are distinguished here by their radial support on the $r$-$z$ plane, which separates the column-confined wave from the one radiating into the ambient. Their dispersion is examined in \S\ref{sec:dispersion}. For each case, the temporal Fourier coefficient $\hat\rho(z,r)$ of the density field is computed at the screech frequency (table~\ref{tab:tones}), using a Hann window over the full 8192-snapshot record. The complex field $\hat\rho(z,r)$ is then interpolated onto a uniform axial grid at each radial station, with the axial extent limited by the local position of the impingement surface. A Fourier transform in $z$ gives $|\hat\rho(k_z,r/D)|$. The upstream-propagating component relevant to feedback-loop closure lies at $k_zD_j<0$, isolated by the streamwise-wavenumber decomposition used to separate upstream- and downstream-travelling waves in screeching jets \citep{edgington2018upstream,edgington2021waves}. Figure~\ref{fig:kx_zd} shows this upstream half of the six-panel wavenumber map, with each panel normalised by its own upstream-half maximum.

\textit{Behaviour on the axis.}\; A cylindrical jet can support the upstream disturbance as one of its intrinsic neutral waves, organised radially by the azimuthal order $m$ \citep{tam1989instability,tam1990theoretical,morris1976spatial}. In figure~\ref{fig:kx_zd} the six cases separate into these two families by their behaviour on the axis alone. Measured against the column maximum at the same wavenumber, the centreline amplitude has a median of $0.53$ and $0.63$ for the axisymmetric ($m=0$) cases Flat[L2.1] and Concave[$\sigma$0.4], against $0.06$-$0.20$ for the four helical ($m=1$) cases. This contrast follows from smoothness: a disturbance of azimuthal order $m$ must scale as $r^{|m|}$ as $r\to0$, so every $m\neq0$ component vanishes on the axis, whereas the $m=0$ component need not. The helical cases accordingly carry a null with two lobes on either side, the $J_1$-type projection of the mode identified from the exit-plane phase field (\S\ref{sec:modes}); here $J_m$ is the Bessel function of the first kind of azimuthal order $m$, which sets the vortex-sheet radial eigenfunction (Appendix~\ref{app:vortex_sheet}). The axisymmetric cases carry none, as an $m=0$ wave permits.

\textit{Radial structure.}\; Away from the axis the two families are no longer distinct. Both peak off the axis, over the overlapping range $|r|/D \approx 0.1$-$0.37$, and decay towards the lip line. In every case the energetic part of the upstream field stays within that line, which is consistent with a wave confined to the jet column rather than a free-field acoustic wave \citep{tam1990theoretical,bogey2017feedback,edgington2018upstream}. The vortex-sheet model predicts no such displacement of the maxima off the axis. It may follow instead from the shock structure of the present jet, which is not ideally expanded; the triple point, the internal shear layer and the recirculation region all fall within the axial window of the transform.

The two families do differ in the shape of this off-axis content. The helical panels carry two `petal-like' structures in the spectra, one of which is outlined in figure~\ref{fig:kx_zd}(d). They broaden with radius, from $|k_zD_j| \approx 5$-$8$ at $|r|/D = 0.15$ to $12$ and beyond by $|r|/D = 0.35$, and fade before the axis is reached. The axisymmetric panels instead carry `streaks' at nearly constant radius, centred near $|r|/D \approx 0.25$ and spanning a comparable range of $k_z$. One reading of the difference, offered as a hypothesis, traces it to where the upstream field is generated. The axisymmetric cases sustain a strong axial Mach-disk pulsation at the tone (\S\ref{sec:machstem}). Such a source is compact in $z$ and centred on the axis, and would imprint a broad band of $k_z$ at every radius it reaches. The helical cases have no tonal Mach-disk motion, and are instead driven from the shear layer, which would leave the content off-axis and tapering. Neither shape is reproduced by the vortex-sheet model, since it admits only discrete neutral modes, each carrying a single axial wavenumber and a fixed radial eigenfunction, and contains neither the shock cells nor the Mach disk that shape the measured field. A weaker, radially extended component nevertheless coexists with all of this; isolating it requires selecting waves that travel at the ambient sound speed, as considered next.

\begin{figure}
\centering
\includegraphics[width=\textwidth]{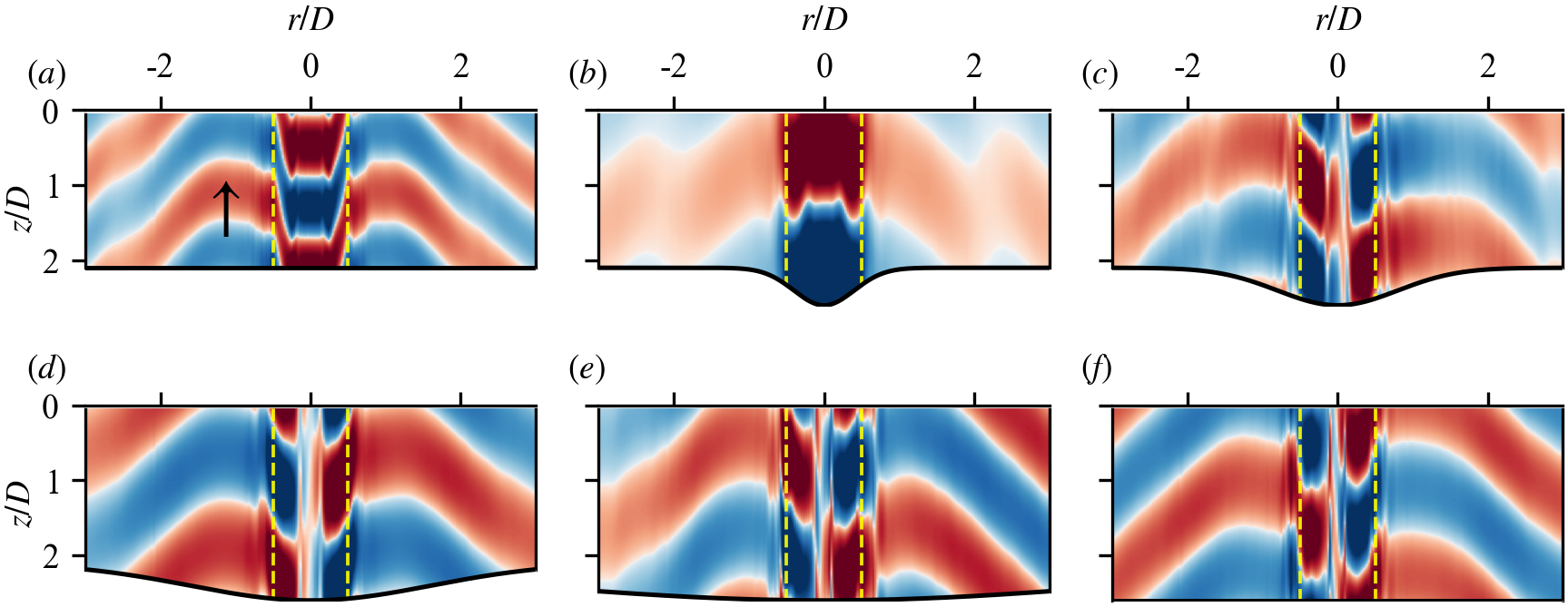}
\caption{Upstream-propagating density fluctuations travelling at the ambient sound speed ($c_\phi = -a_\infty$) for the six configurations, each band-limited to its primary tone $St_1$ (table~\ref{tab:tones}): (a) Flat[L2.1], (b) Concave[$\sigma$0.4], (c) Concave[$\sigma$0.8], (d) Concave[$\sigma$1.6], (e) Concave[$\sigma$4.0] and (f) Flat[L2.6]. Black solid lines mark the impingement surfaces and yellow dashed lines the lip line at $|r|/D = 0.5$; the arrow in panel~(a) indicates the direction of upstream phase propagation. Each case is normalised by the envelope of its own reconstruction. An animated rendering over one screech-tone period is included as Movie 2.}
\label{fig:kx_snap_up}
\end{figure}

\textit{Internal and external contributions.}\; The upstream-propagating component is now transformed back to physical space on the meridional $r$-$z$ plane and shown in figure~\ref{fig:kx_snap_up}. The upstream half of the wavenumber spectrum is first restricted to a band centred on the acoustic wavenumber, $k_a=2\pi f_1/a_\infty$. Across the six cases, this wavenumber lies within $|k_zD_j|=2.2$-$4.3$, and a band of half-width $0.2$-$0.4$ is used. Waves within this band propagate upstream at the ambient sound speed, so the restriction separates the externally radiating acoustic contribution from the slower, column-trapped content \citep{towne2017acoustic,unnikrishnan2016acoustic,martini2020resolvent}. Transforming the restricted coefficient back in $z$ yields the complex upstream density field $\hat\rho^{-}(z,r)$, whose real part is plotted for all six configurations. The corresponding phase evolution, $\mathrm{Re}\,[\hat\rho^{-}(z,r)\exp(-\mathrm{i}2\pi f_1 t)]$, over one screech-tone period, $T_1=1/f_1$, is provided as Movie 2.

Despite propagating at the ambient sound speed, the field in figure~\ref{fig:kx_snap_up} is strongest within the jet core, $|r|/D<0.5$, in all six cases. The acoustic-speed content is therefore concentrated primarily within the jet column rather than in the surrounding ambient. Its radial organisation follows the Powell classification: the field is symmetric and axis-peaked for the mode-A cases, and two-lobed for the mode-C cases. A weaker component extends beyond $|r|/D=1$ and curves outward as it propagates upstream from the impingement region. This component corresponds to the externally radiating wave, which has been observed in impinging jets to propagate upstream outside the plume and perturb the nozzle lip \citep{edgington2019aeroacoustic}. The internal wave dominates the external contribution most strongly for Concave[$\sigma$0.4]. In this case, the narrow indentation lies almost entirely within $|r|/D<0.5$, so the wall interacts with the returning disturbance inside the jet column rather than primarily through the reflected ambient field. Consequently, this case has the largest internal-to-external amplitude ratio among the six configurations. The column-confined wave is therefore the dominant upstream component. Whether it is a genuine guided jet mode, and whether it determines the tone, is examined through the dispersion comparison in \S\ref{sec:dispersion}. Both the internal and external components amplify as the indentation narrows, in the proportions quantified in \S\ref{sec:why}.

\begin{figure}
\centering
\includegraphics[width=\textwidth]{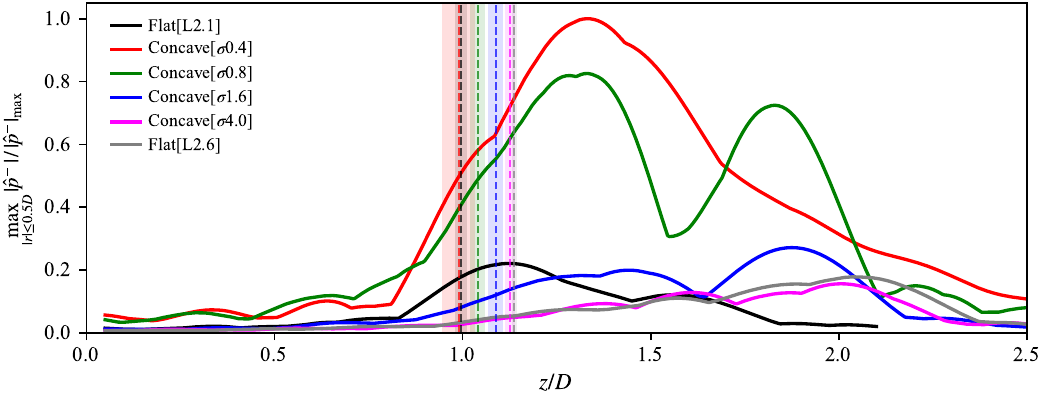}
\caption{Column-region envelope of the upstream-propagating pressure field, $\max_{|r|\le 0.5D}|\hat p^{-}|$, along the axial coordinate for the six configurations, all normalised by the overall maximum so that amplitudes are directly comparable. The field $\hat p^{-}$ is obtained from the upstream-propagating half ($k_zD_j<0$) of the pressure projected onto the primary tone. Dashed vertical lines mark the mean Mach-stem location of each case in the matching colour, the shaded band spanning one standard deviation of its axial excursion.}
\label{fig:upstream_amp}
\end{figure}

\textit{Dependence on wall geometry.}\; The upstream-wave amplitude varies strongly with wall geometry, providing a direct test of the source-transfer attribution in \S\ref{sec:why}. Figure~\ref{fig:upstream_amp} compares the column-region envelope of the upstream pressure field, $\max_{|r|\le 0.5D}|\hat p^{-}|$, along the axial direction for all six configurations on a common amplitude scale. The full upstream-propagating half, $k_zD_j<0$, is retained rather than only the $c_\phi=-a_\infty$ band used in figure~\ref{fig:kx_snap_up}, because the upstream tonal field is dispersive and most of its amplitude resides in the slower column-trapped component. The temporal projection is also taken over a narrow Strouhal band about each $St_1$, rather than at a single spectral line, so that tonal energy migrating between neighbouring bins over the record (\S\ref{sec:nearfield}) is retained.

Among the concave walls, the two narrowest indentations, Concave[$\sigma$0.4] and Concave[$\sigma$0.8], produce both the largest screech amplification and the strongest upstream-propagating waves. As the indentation is widened, the column-region amplitude decreases by $11$ to $16$~dB. In all cases, the envelope maximum lies downstream of the axial range swept by the Mach disk, by more than six standard deviations of its motion, indicating that the peak is not a direct consequence of shock displacement. Across the concave family, the increase in screech amplitude is therefore accompanied by a comparable increase in the upstream-propagating field, consistent with the loop-gain decomposition in \S\ref{sec:why}. Flat[L2.1] does not follow this trend. Its primary tone is the third strongest of the six cases, yet its column-region envelope is among the weakest and remains below that of Concave[$\sigma$1.6]. This is also the case that departs from the source-transfer balance in \S\ref{sec:why}, and the only one whose amplification relative to Flat[L2.6] arises from a change in standoff distance rather than wall curvature. The correspondence between tone amplitude and upstream-wave amplitude is therefore established for the concave family, but not for changes in impingement distance.

\subsection{Dispersion analysis of upstream-propagating waves}
\label{sec:dispersion}
The wavenumber spectra of \S\ref{sec:waves_pod} fix the radial organisation of the upstream-propagating waves but not their dispersion. To test whether the upstream tonal energy coincides with a GJM of the equivalent ideally expanded jet, the frequency-wavenumber spectrum $|S(k_z,\omega)|$ of the density fluctuation is computed on the cylindrical surface $r = 0.35\,D$ inside the plume. This radius is chosen because the helical modes vanish on the axis: it lies close to the maximum of the $H_1$ eigenfunction, at $r/D = 0.30$, and the measured field retains at least half of its peak column amplitude there in every case. It is projected onto each case's dominant azimuthal Powell mode: $m = 0$ for the axisymmetric cases Flat[L2.1] and Concave[$\sigma$0.4], and $m = 1$ for the helical cases Concave[$\sigma$0.8], Concave[$\sigma$1.6], Concave[$\sigma$4.0] and Flat[L2.6]. Figure~\ref{fig:dispersion} presents the resulting spectra, overlaid with the vortex-sheet dispersion branches of Appendix~\ref{app:vortex_sheet}: the axisymmetric $A_n$ branches on the $m=0$ panels and the helical $H_n$ branches on the $m=1$ panels. Also marked are the acoustic-speed line $k_z = -\omega/a_\infty$ equation~(\ref{eq:app_acline}) and the analytic lower limits $St_{\min,n}$ equations~(\ref{eq:app_axisym_min})-(\ref{eq:app_helical_min}). The branches and limits are evaluated at the present $M_j = 1.56$.

\begin{figure}
    \centering
    \includegraphics[width=\textwidth]{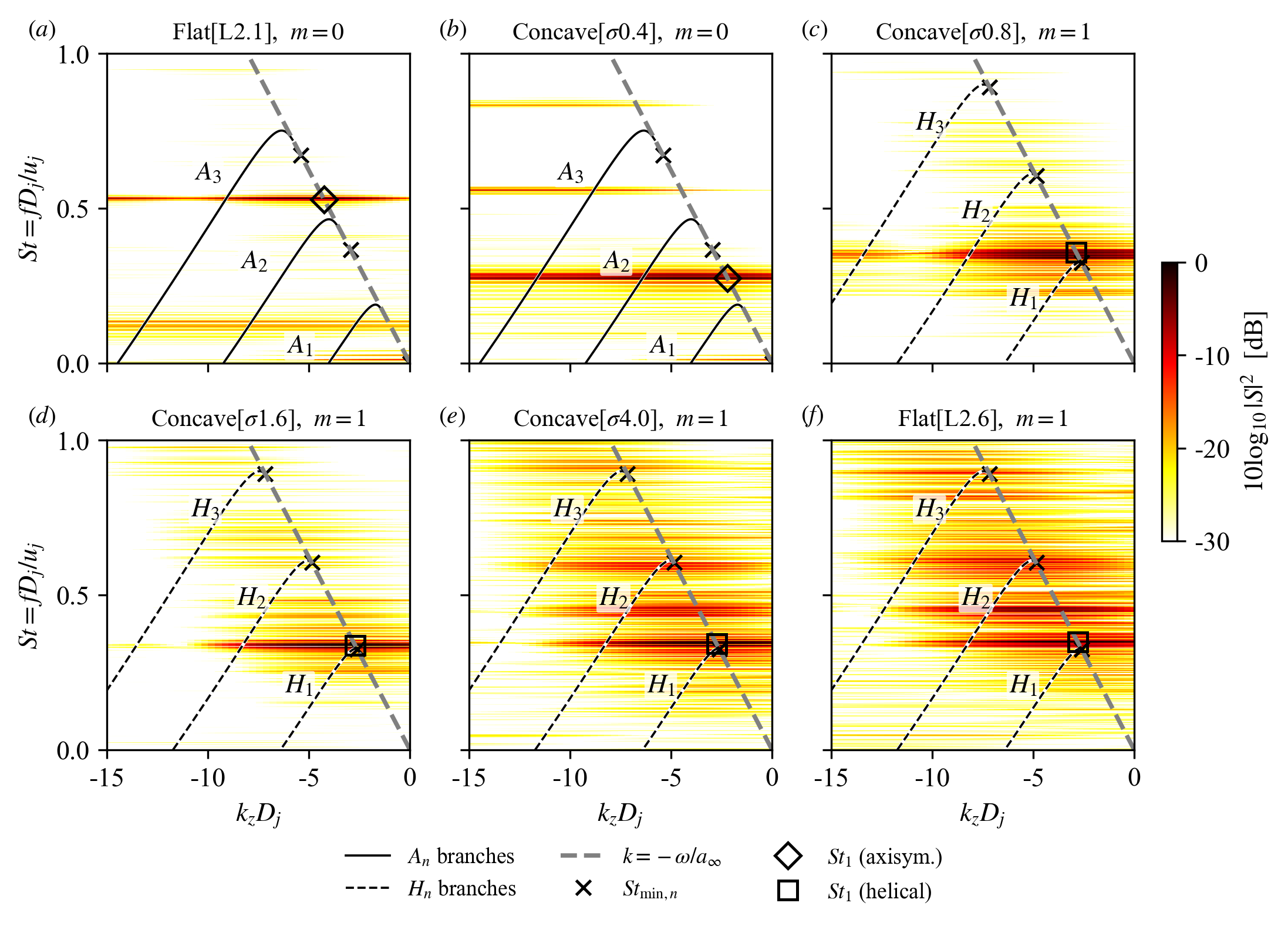}
    \caption{Frequency-wavenumber spectrum of the upstream density field on the $r = 0.35\,D$ cylinder, projected onto each case's dominant azimuthal mode, with the vortex-sheet dispersion branches overlaid. Colour: $10\log_{10}|S(k_z D_j, St)|^2$ in dB. Solid black: axisymmetric $A_n$ branches; dashed black: helical $H_n$ branches (Appendix~\ref{app:vortex_sheet}, equation~\ref{eq:app_disp}); grey dashed: the acoustic-speed line $k_z = -\omega/a_\infty$ equation~(\ref{eq:app_acline}). Markers: $\times$ analytic lower limits $St_{\min,n}$ equations~(\ref{eq:app_axisym_min})-(\ref{eq:app_helical_min}); $\diamond$ axisymmetric primary tone $St_1$; $\square$ helical primary tone. Panels: (a) Flat[L2.1] and (b) Concave[$\sigma$0.4], dominant mode $m=0$; (c) Concave[$\sigma$0.8], (d) Concave[$\sigma$1.6], (e) Concave[$\sigma$4.0] and (f) Flat[L2.6], dominant mode $m=1$.}
    \label{fig:dispersion}
\end{figure}

\begin{figure}
    \centering
    \includegraphics[width=\textwidth]{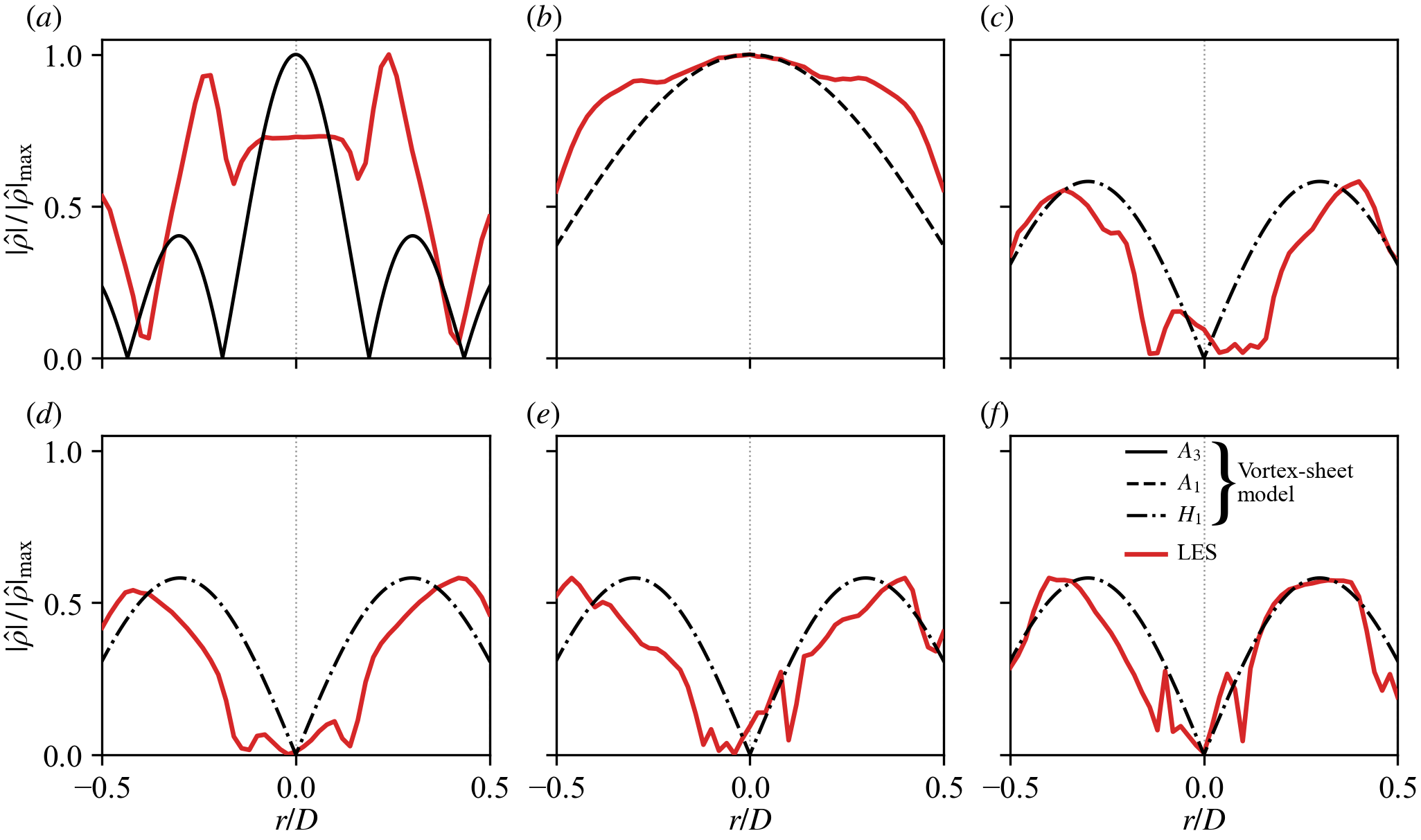}
    \caption{Radial eigenfunctions of the upstream $c_\phi = -a_\infty$ density component over the jet core, $r/D \in [-0.5, 0.5]$, each at its primary tone $St_1$ (table~\ref{tab:tones}): (a) Flat[L2.1], (b) Concave[$\sigma$0.4], (c) Concave[$\sigma$0.8], (d) Concave[$\sigma$1.6], (e) Concave[$\sigma$4.0] and (f) Flat[L2.6]. Red: phase-matched LES amplitude from the band-limited field of figure~\ref{fig:kx_snap_up}, normalised for the helical cases to the off-axis $|J_1|$ peak. Black: vortex-sheet eigenfunction, equation~(\ref{eq:app_eigfun}), of the branch of closest order, $A_3$ for (a), $A_1$ for (b) and $H_1$ for (c)-(f), each evaluated at its own lower-limit corner wavenumber rather than at the measured tone. The two frequencies agree to within $10\%$ for the helical cases, but not for the axisymmetric ones, which coincide with no neutral $A_n$ branch (\S\ref{sec:dispersion}).}
    \label{fig:eigenfunction}
\end{figure}

In each case, the tonal energy forms a horizontal band at the primary Strouhal number $St_1$ and intersects the acoustic-speed line, indicating an upstream-propagating tonal wave with phase speed close to $-a_\infty$. The two mode families differ in their relation to the guided-jet-mode branches. For the helical cases [figure~\ref{fig:dispersion}(c-f)], the tonal band meets the $H_1$ branch at its lower-limit corner on the acoustic-speed line. The $St_1$ marker coincides with the point at which the $H_1$ curve meets the acoustic-speed line, where its phase speed is $c_\phi=-a_\infty$, so the helical tones lie at the $H_1$ lower limit. This agreement holds for all four helical cases, from the narrow indentation Concave[$\sigma$0.8] to the flat plate Flat[L2.6], whose tones span only $St_1=0.336$-$0.357$. The $H_1$ lower-limit selection is therefore recovered for both concave and flat walls, consistent with a guided jet mode set primarily by the shear-layer profile of the equivalent ideally expanded jet rather than by the downstream geometry. Similar lower-limit locking of an upstream-propagating guided-jet-mode branch has been reported for axisymmetric screech modes of underexpanded free jets by \citet{li2020acoustic} and for underexpanded jets impinging on an inclined plate up to $M_j=1.56$ by \citet{li2023acoustic}. For the axisymmetric cases [figure~\ref{fig:dispersion}(a,b)], the tonal band does not coincide with a neutral $A_n$ branch: $St_1=0.529$ for Flat[L2.1] and $St_1=0.275$ for Concave[$\sigma$0.4] fall between successive axisymmetric lower limits on the acoustic-speed line. As shown next, however, the LES upstream field at these tones still retains a column-confined radial distribution that broadly resembles a low-order $J_0$ eigenfunction near the axis.

The radial structure of the upstream wave confirms that its dominant component remains confined to the jet column. Figure~\ref{fig:eigenfunction} compares, within the jet core $|r|/D \le 0.5$, the radial amplitude of the $c_\phi=-a_\infty$ component isolated in figure~\ref{fig:kx_snap_up} with the vortex-sheet eigenfunction equation~(\ref{eq:app_eigfun}) of the branch nearest each tone. The helical profiles reproduce the $J_1$ organisation of the $H_1$ branch, with an on-axis null and off-axis lobes. The LES lobes peak slightly farther from the axis, near $|r|/D \approx 0.4$, than the $H_1$ eigenfunction maximum at $r/D \approx 0.27$, and retain finite amplitude at the lip line. The axisymmetric profiles are also column-confined, with amplitude concentrated near the jet axis as in low-order $J_0$ eigenfunctions, but their detailed radial structure is not captured. The Flat[L2.1] profile is broadly peaked across the core, with only a shallow central depression rather than the two clear internal nodes of the $A_3$ branch, while the node-free Concave[$\sigma$0.4] profile is substantially broader than $A_1$.

The LES profiles are largest within the jet core in every case, but quantitative agreement with the vortex-sheet eigenfunctions is limited to the helical cases. For these cases, the two independent diagnostics agree closely: the observed tone coincides with the $H_1$ lower-limit corner, and the radial profile reproduces the $J_1$ eigenfunction, including its on-axis null and off-axis lobes. This agreement, obtained consistently across all four helical configurations, provides strong evidence that the guided jet mode governs the upstream leg of the helical feedback loop. For the axisymmetric cases, the evidence is less conclusive. The wave is column-confined and qualitatively $J_0$-like near the axis, but its detailed radial profile is not reproduced and the observed tone does not coincide with any $A_n$ branch. Thus, the vortex-sheet model captures neither the eigenfunction in detail nor the selected frequency for the axisymmetric modes. Whether a guided jet mode sets the axisymmetric tone therefore remains unresolved; the discrepancy may reflect the departure of the present jet from ideal expansion, or a mechanism not represented by the inviscid vortex-sheet model. In all six cases, the externally radiating wave extending beyond $|r|/D=1$ in figure~\ref{fig:kx_snap_up} remains a weaker co-participant at the same phase speed, $c_\phi=-a_\infty$, and amplifies with the curvature alongside the column-confined component (\S\ref{sec:why}).

\subsection{Frequency-selection mechanisms}
\label{sec:feedback}

\begin{figure}
    \centering
    \includegraphics[width=\textwidth]{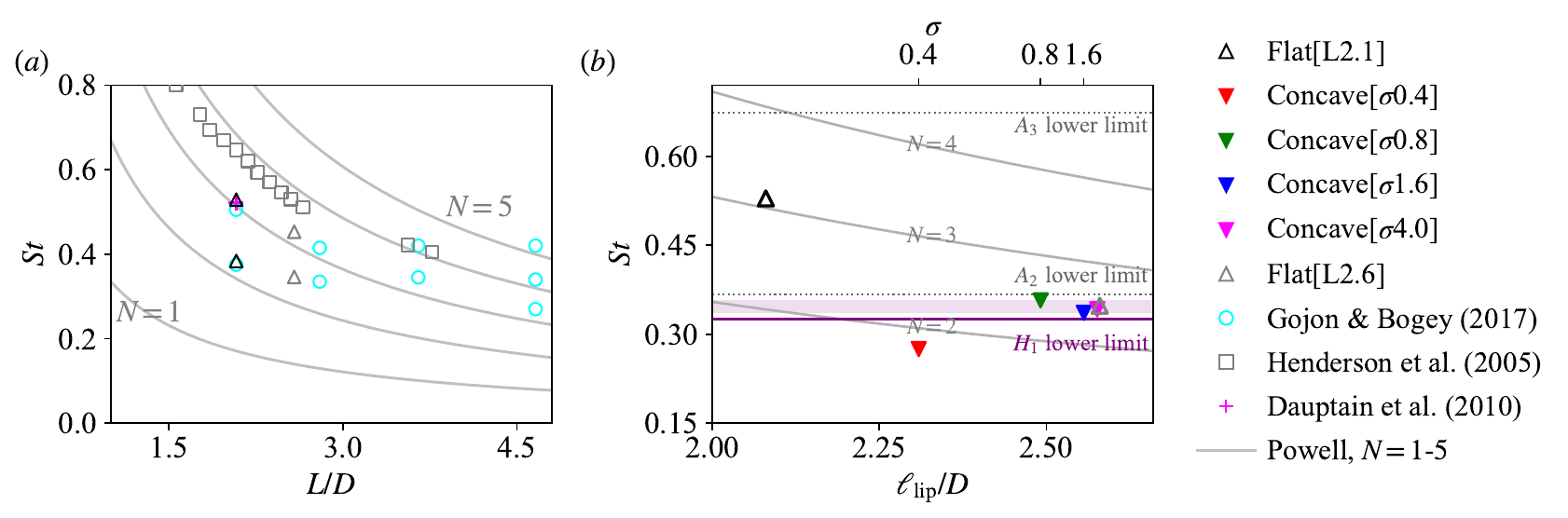}
    \caption{Frequency selection of the dominant screech tone across the six configurations. (a) Screech-tone Strouhal numbers against the nozzle-to-wall distance $L/D$; for the two flat-wall configurations both the primary and the secondary tone are plotted. The gray lines are the Powell feedback-mode curves for $N=1$-$5$ equation~(\ref{eqn:powell_feedback}). The present flat-wall tones are consistent with the impinging-jet measurements of \citet{henderson2005experimental}, \citet{dauptain2010large} and \citet{gojon2017flow}, and follow the mode-staging ladder traced by the gray lines. (b) Frequency-selection map against the lip-line feedback length $\ell_{\rm lip}/D$ (lower axis; the indentation spread $\sigma$ of the concave cases is on the aligned upper axis). Grey curves are the Powell feedback ladder $N=2$-$4$ equation~(\ref{eqn:powell_feedback}). The horizontal lines are the vortex-sheet GJM lower limits at $M_j=1.56$: the helical $H_1$ limit (purple line) and the axisymmetric $A_2$, $A_3$ limits (dotted). The purple shading spans the four measured helical tones and shows how closely they cluster above the $H_1$ limit.}
    \label{fig:tones}
\end{figure}

The convection velocities from \S\ref{sec:Uc} and the upstream-wave characteristics from \S\ref{sec:dispersion} are now combined into a phase criterion for screech resonance, following Powell's feedback interpretation \citep{powell1953mechanism,edgington2019aeroacoustic}. Resonance requires the phase accumulated over one complete loop, downstream along the shear layer and then upstream to the nozzle lip, to be an integer multiple of $2\pi$,
\begin{equation}
(k_d - k_u)\,\ell = 2\pi N ,
\label{eqn:phase_criterion}
\end{equation}
where $k_d = \omega/u_c > 0$ is the wavenumber of the downstream-convecting Kelvin-Helmholtz wavepacket, $k_u < 0$ is the wavenumber of the upstream-travelling wave, $\ell$ is the loop length, and $N$ is an integer. If the upstream leg is treated as a free wave travelling at the ambient sound speed, then $k_u = -\omega/a_\infty$. With $\omega = 2\pi f$, equation~(\ref{eqn:phase_criterion}) reduces to the classical Powell form,
\begin{equation}
\frac{N}{f} = \frac{\ell}{u_c} + \frac{\ell}{a_\infty} ,
\label{eqn:powell_feedback}
\end{equation}
in which the loop period is the sum of the downstream convection time and the upstream return time. The mode number $N$ counts the cells in the standing-wave pattern formed by the downstream and upstream waves over the loop length. It is $N=3$ for Flat[L2.1] and $N=2$ for the other cases, with the measured $u_c$ entering directly. The continuous feedback-mode ladder in figure~\ref{fig:tones} is obtained by evaluating equation~(\ref{eqn:powell_feedback}) over the range of loop lengths, using the empirical convection-velocity relation in equation~(\ref{eqn:gb_uc}) to provide $u_c(L)$. The case-by-case predictions in table~\ref{tab:st1_predictions}, by contrast, use the measured $u_c$. Equation~(\ref{eqn:phase_criterion}) also makes the selection mechanism clear: the resonant frequency depends on the upstream wavenumber $k_u$, so the loop length $\ell$ alone is decisive only when the upstream leg is non-dispersive. A guided jet mode instead propagates upstream only within a neutral frequency band and has its own relation $k_u(\omega)$, which can determine the frequency independently of $\ell$. The vortex-sheet dispersion analysis in \S\ref{sec:dispersion} therefore adds the requirement that the upstream leg correspond to a propagating guided jet mode.

For a curved wall, the loop length is not uniquely defined. Table~\ref{tab:st1_predictions} compares two choices: the lip-line distance $\ell_{\rm lip}$, given by equation~(\ref{eqn:ell_lip}), which corresponds to the conventional impingement distance, and the footprint-mean distance $L_{\rm avg}$ over the base of the jet column, $r\le 0.5D$, given by equation~(\ref{eqn:L_avg}). The latter is motivated by the upstream field in \S\ref{sec:waves_pod}, whose amplitude is concentrated within the column, $|r|/D<0.5$. The returning wave therefore samples the wall across the full column footprint rather than only at the lip line. For a concave indentation, which is deepest on the axis, this effective return path is longer than $\ell_{\rm lip}$; for flat walls, the two lengths coincide.

\begin{table}
\begin{center}
\def~{\hphantom{0}}
\small
\begin{tabular}{lccccccc}
 & & & & & \multicolumn{2}{c}{Powell $St_1$} & GJM $St_1$ \\
\cmidrule(lr){6-7}\cmidrule(lr){8-8}
Case & Mode & $N$ & $u_c/u_j$ & $St_1$ (LES) & $\ell_{\rm lip}$ & $L_{\rm avg}$ & $H_1$ \\[4pt]
Flat[L2.1]            & A & 3 & 0.561 & 0.529 & 0.520\,($-1.7$)  & 0.520\,($-1.7$)  & -                \\
Concave[$\sigma$0.4]  & A & 2 & 0.599 & 0.275 & 0.324\,($+17.9$) & 0.309\,($+12.2$) & -                \\
Concave[$\sigma$0.8]  & C & 2 & 0.591 & 0.357 & 0.298\,($-16.4$) & 0.293\,($-17.9$) & 0.325\,($-8.9$)   \\
Concave[$\sigma$1.6]  & C & 2 & 0.594 & 0.336 & 0.292\,($-13.2$) & 0.290\,($-13.6$) & 0.325\,($-3.2$)   \\
Concave[$\sigma$4.0]  & C & 2 & 0.588 & 0.342 & 0.288\,($-15.9$) & 0.287\,($-16.0$) & 0.325\,($-4.9$)   \\
Flat[L2.6]            & C & 2 & 0.585 & 0.348 & 0.286\,($-17.7$) & 0.286\,($-17.7$) & 0.325\,($-6.5$)   \\
\end{tabular}
\caption{LES primary screech tone $St_1$ compared with two feedback-loop estimates and the GJM prediction. The Powell estimate equation~(\ref{eqn:powell_feedback}) is evaluated with both the lip-line distance $\ell_{\rm lip}$ equation~(\ref{eqn:ell_lip}) and the footprint-mean distance $L_{\rm avg}$ over $r\le 0.5\,D$ equation~(\ref{eqn:L_avg}), using the case-specific convection velocity $u_c/u_j$ from \S\ref{sec:Uc} and the standing-wave mode number $N$ from \S\ref{sec:modes}. The GJM column is the lower-limit Strouhal number of the helical $H_1$ branch of the vortex-sheet model (Appendix~\ref{app:vortex_sheet}), which is independent of the wall geometry. Values in parentheses are the percentage deviation from the LES tone. A dash indicates that no neutral $A_n$ branch coincides with the axisymmetric tones.}
\label{tab:st1_predictions}
\end{center}
\end{table}

For the axisymmetric configurations, the loop-length interpretation is adequate (table~\ref{tab:st1_predictions}, figure~\ref{fig:tones}). The Flat[L2.1] tone is recovered to within $2\%$ using either loop length. For Concave[$\sigma$0.4], the lip-line length overpredicts the frequency by $18\%$, while the footprint-mean length reduces the overprediction to $12\%$. This improvement is consistent with the deeper on-axis return path sampled by the column-confined upstream wave. Part of the remaining overprediction can be attributed to the convection velocity. The value $u_c=0.599\,u_j$ used for this case is the highest among the six configurations and is also the least reliable, because no coherent wavepacket could be tracked beyond $z/D\approx1.1$ (\S\ref{sec:Uc}). Matching the LES tone would require $u_c\approx0.46$-$0.50\,u_j$, which is below the empirical estimate of $\approx0.55\,u_j$ from equation~(\ref{eqn:gb_uc}). The convection velocity therefore explains part, but not all, of the overprediction. Both axisymmetric tones lie on the low-frequency $N$ branch and remain well separated from the helical band. Neither coincides with a neutral $A_n$ branch of the vortex-sheet model, whose nearest lower limits are $St=0.37$ and $0.67$. The guided-mode constraint therefore does not select these tones. For these two cases, the axisymmetric resonance is described primarily by Powell's classical loop-length criterion.

The helical configurations behave differently. The loop-length model underpredicts their tones by $13$-$18\%$ for either choice of length, and the predicted frequency varies with indentation spread. By contrast, the measured tones are nearly invariant, spanning only $0.336 \le St_1 \le 0.357$ across the four cases. The selection is therefore dispersive rather than geometric. In the frequency-selection diagram of figure~\ref{fig:tones}(b), the four helical tones lie within a narrow strip immediately above the $H_1$ lower-limit line, indicated by the purple shading, and away from the Powell ladder, irrespective of $\ell_{\rm lip}$. This lower limit, $St_1 \simeq 0.325$, occurs where the $H_1$ guided-jet-mode branch meets the acoustic line. At this point, its phase speed equals $-a_\infty$, and the mode ceases to be radially evanescent in the ambient (Appendix~\ref{app:vortex_sheet}). It defines the lower edge of the frequency band over which the $H_1$ mode propagates upstream. The helical resonance therefore locks to this band edge rather than to the loop length. The radial distribution of the LES upstream field matches the $H_1$ eigenfunction (\S\ref{sec:dispersion}), confirming this identification. Because the lower limit is set by the shear-layer profile of the equivalent ideally expanded jet, it is insensitive to the wall geometry. Accordingly, the helical tone changes little across the concave and flat cases and lies within $3$-$9\%$ of the $H_1$ limit. The largest departure, approximately $9\%$ for the narrowest helical indentation Concave[$\sigma$0.8], may indicate a weak residual sensitivity to wall geometry. Nevertheless, the $H_1$ lower limit remains an accurate predictor across the helical family.

The two mode families are therefore selected by different conditions within the same feedback framework. The phase criterion in equation~(\ref{eqn:phase_criterion}) admits a ladder of loop frequencies determined by $\ell$ and $u_c$, while the vortex-sheet dispersion relation restricts the upstream leg to a propagating guided jet mode within its neutral band. For the helical cases, these conditions coincide near the $H_1$ lower limit. The band-edge frequency then controls the selection, so the tone is governed primarily by shear-layer dispersion and is insensitive to the indentation spread. For the axisymmetric cases, the observed tone follows the loop-length ladder but coincides with no $A_n$ band edge. This behaviour is consistent with loop-length rather than guided-mode selection. Whether the absence of an $A_n$ match results from the departure of the present jet from ideal expansion remains unresolved.

\section{Conclusions}
\label{sec:conclusion}

LES have been performed to determine how the curvature of the impinging wall alters the flow-structure oscillations, surface loading and aeroacoustic resonance of underexpanded round supersonic jets. The operating conditions correspond to an ideally expanded jet Mach number $M_j=1.56$ and Reynolds number $6\times 10^4$. Six geometries were considered: two flat walls, corresponding to the limiting spreads $\sigma\to0$ and $\sigma\to\infty$, and four Gaussian concave surfaces of fixed depth that interpolate between them as $\sigma$ is varied. The curvature therefore enters through a single parameter, the indentation spread. Two questions were addressed: how the near-field acoustics and the unsteady wall loading respond to that spread, and how curvature acts on the screech resonance mechanism through its amplification and upstream closure. The principal findings are as follows.

\begin{enumerate}

\item Wall curvature affects the resonance primarily when the indentation overlaps the jet column. As the indentation narrows toward the radial scale of the jet, the primary tone increases progressively, reaching $\approx 23$~dB for the narrowest concave wall relative to the flat-wall reference at $L/D=2.6$, and $\approx 19$~dB relative to the widest concave wall considered here. In contrast, an indentation extending well into the entrained ambient produces a near-field spectrum close to that of the flat-wall case. The dominant tone in each configuration is carried by a single azimuthal mode, either axisymmetric or helical, and the Mach-disk motion follows this symmetry across all six cases: the two axisymmetric cases show a strong, periodic axial pulsation of the Mach disk, whereas the four helical cases show only weak, broadband motion. A toroidal disturbance reaches the Mach disk in phase around the azimuth and a helical one does not, which accounts for the correspondence.

\item The curvature-induced increase in screech amplitude arises from two complementary mechanisms. A Powell-Tam source-transfer budget shows that part of the amplification is associated with increased source strength at the Mach disk, while the remainder appears as an upstream transfer/receptivity contribution. Direct measurements of the upstream-propagating wave show that it amplifies together with the tone in both the jet column and the ambient, and that the column carries the larger amplitude in every configuration, by $14$ to $30$~dB. The column is also the channel that responds most to the curvature, gaining $15.8$~dB more than the ambient at the narrowest indentation, whose curvature lies almost entirely beneath the jet column, against $1.3$~dB or less at the wider ones. Geometric focusing therefore acts through the plume and through the ambient in proportions set by how far the indentation reaches beyond the jet column.

\item The symmetry of the screech mode determines the unsteady surface loading through the azimuthal orthogonality of the wall pressure, equation~(\ref{eqn:azim_projection}). Only the $m=0$ pressure mode contributes to the net normal force, while only the $m=\pm1$ modes contribute to the transverse force and in-plane bending moment. Axisymmetric screech therefore produces a tonal normal force with negligible transverse loading. By contrast, helical screech makes the normal force broadband and concentrates the tonal response in a bending moment that precesses about the jet axis. This moment follows an elliptical, rather than circular, orbit, with its eccentricity indicating the imbalance between the two counter-rotating helical components. The mode symmetry therefore determines the direction of the fatigue-relevant unsteady load: axial for axisymmetric screech and rotating in-plane for helical screech. The largest bending moments occur at intermediate indentation spread, rather than at the strongest curvature.

\item The radial organisation of the upstream-propagating wave distinguishes the two resonance types. In the helical cases, the disturbance is concentrated within the jet column and displays the off-axis, double-lobed radial pattern of the $H_1$ guided-jet-mode eigenfunction predicted by the vortex-sheet model. The measured screech frequencies also lie just above the lower-frequency limit of the corresponding $H_1$ branch. Together, these findings provide strong evidence that a guided jet mode closes the upstream leg of the helical feedback loop. The axisymmetric cases also exhibit an upstream-propagating wave that remains largely confined to the jet column, with fluctuation amplitudes concentrated near the jet axis. However, neither the radial organisation nor the measured frequency agrees with any $A_n$ eigenfunction of the vortex-sheet model.

\item The two tone families observed across the six configurations follow different selection conditions. The four helical tones lie within $3$-$9\%$ of the lower-frequency edge of the $H_1$ branch and span only $0.336 \le St_1 \le 0.357$, despite the variation in loop length. The same selection occurs for both concave and flat walls, consistent with a guided jet mode governed primarily by the shear-layer profile of the equivalent ideally expanded jet rather than by the downstream geometry. By contrast, the two axisymmetric tones follow the loop-length ladder predicted by Powell's phase criterion and coincide with no neutral $A_n$ branch of the vortex-sheet model. Whether this mismatch results from the departure of the present jet from ideal expansion remains unresolved.

\end{enumerate}

\vspace{10pt}
\noindent \textbf{Movies.} \label{SupMat}
\begin{enumerate}
    \item \href{https://youtu.be/DEZ_qcgXfNE}{\textbf{Movie1.mp4}} - Instantaneous density and velocity dilatation fields on the meridional $r$-$z$ plane
    \item \href{https://youtu.be/sm9bdnvc1Bo}{\textbf{Movie2.mp4}} - Upstream-propagating density fluctuations travelling at the ambient sound speed, band-limited to the primary screech tone and animated over one screech-tone period (figure~\ref{fig:kx_snap_up})
\end{enumerate}

\vspace{10pt}
\noindent \textbf{Acknowledgements.} Computational resources were provided by the Division of Computing and Information Systems, Technion - Israel Institute of Technology.

\vspace{10pt}
\noindent \textbf{Declaration of interest.} The authors report no conflict of interest.

\vspace{10pt}
\noindent \textbf{Declaration of AI assistance.} AI tools (ChatGPT 5.5) were used solely to improve the clarity and grammar of the manuscript text. All content was reviewed and verified by the authors, who take full responsibility for the final text.

\appendix

\section{Vortex-sheet model of the upstream-propagating GJMs}\label{app:vortex_sheet}

The closed-form expressions used in \S\ref{sec:dispersion} to overlay the GJM dispersion branches on the LES wavenumber-frequency spectra, and to compare with the radial structure of the LES upstream field, are those of the cylindrical vortex-sheet model of \citet{tam1989instability,tam1990theoretical}. They are restated here for completeness and ease of reproduction.

\subsection{Model and eigenfunctions}

Following \citet{tam1989instability,tam1990theoretical}, the ideally expanded jet is represented as a uniform cylindrical stream of radius $r_0 = D_j/2$, axial velocity $u_j$, sound speed $a_j$ and density $\rho_j$, separated by an infinitesimally thin vortex sheet at $r = r_0$ from a quiescent ambient of sound speed $a_\infty$ and density $\rho_\infty$. Continuity of static pressure across the sheet gives $\rho_j a_j^2 = \rho_\infty a_\infty^2$. Linearising the compressible Euler equations about this base state and writing each perturbation as a normal mode,
\begin{equation}
q'(z,r,\theta,t) = \mathrm{Re}\!\left[\hat q(r)\,\exp\!\big(\mathrm{i}(k_z z + m\theta - \omega t)\big)\right],
\label{eq:app_mode}
\end{equation}
with real axial wavenumber $k_z$, integer azimuthal order $m$ ($m=0$ axisymmetric, $m=1$ helical) and real angular frequency $\omega$, the latter because the modes of interest are neutral. The pressure amplitude $\hat p(r)$ then satisfies Bessel's equation of order $m$ in each uniform region,
\begin{equation}
\frac{\mathrm{d}^2\hat p}{\mathrm{d}r^2} + \frac{1}{r}\frac{\mathrm{d}\hat p}{\mathrm{d}r} + \left(\kappa^2 - \frac{m^2}{r^2}\right)\hat p = 0,
\end{equation}
in which the radial wavenumber $\kappa$ takes the values
\begin{equation}
\kappa_j^2 = \frac{(\omega - k_z u_j)^2}{a_j^2} - k_z^2 \quad (r<r_0), \qquad
\kappa_\infty^2 = \frac{\omega^2}{a_\infty^2} - k_z^2 \quad (r>r_0).
\end{equation}
The upstream GJMs are oscillatory inside the jet ($\kappa_j^2>0$) and evanescent in the ambient ($\kappa_\infty^2<0$). Requiring regularity on the axis and decay as $r\to\infty$ selects
\begin{equation}
\hat p(r) =
\begin{cases}
A\,J_m\!\big(|\kappa_j|\,r\big), & r \le r_0,\\[4pt]
B\,K_m\!\big(|\kappa_\infty|\,r\big), & r \ge r_0,
\end{cases}
\label{eq:app_eigfun}
\end{equation}
where $J_m$ is the Bessel function of the first kind and $K_m$ the modified Bessel function of the second kind. Equation~\eqref{eq:app_eigfun} is the closed-form eigenfunction. Inside the jet the radial shape is the oscillatory $J_m$, which peaks on the axis for the axisymmetric modes ($m=0$) and has an on-axis null with off-axis lobes for the helical modes ($m=1$). Outside the jet it decays monotonically as $K_m$. This $J_m$/$K_m$ partition is the prediction compared with the LES radial eigenfunctions extracted at the screech tones.

\subsection{Dispersion relation}

Introducing the phase Mach number relative to the ambient, $C = \omega/(k_z a_\infty)$, the radial-wavenumber factors
\begin{equation}
\xi_+ = \left|C^2 - 1\right|^{1/2}, \qquad
\xi_- = \left|\Big(\tfrac{a_\infty}{a_j}\,C - M_j\Big)^{2} - 1\right|^{1/2}, \qquad
\alpha = k_z r_0 = \tfrac{1}{2}\,k_z D_j,
\end{equation}
satisfy $|\kappa_\infty|\,r_0 = |\xi_+\alpha|$ and $|\kappa_j|\,r_0 = |\xi_-\alpha|$. Continuity of pressure and of the radial displacement of the sheet across $r=r_0$, the latter expressed as $\hat p_j'(r_0)/[\rho_j(\omega-k_z u_j)^2] = \hat p_\infty'(r_0)/(\rho_\infty\,\omega^2)$, close the system. The resulting dispersion relation,
\begin{equation}
\xi_+\,\frac{J_m(|\xi_-\alpha|)\,\big[K_{m-1}(|\xi_+\alpha|) + K_{m+1}(|\xi_+\alpha|)\big]}{K_m(|\xi_+\alpha|)}
+ \frac{C^2\,\xi_-}{\big(\tfrac{a_\infty}{a_j}C - M_j\big)^{2}}\,\big[J_{m-1}(|\xi_-\alpha|) - J_{m+1}(|\xi_-\alpha|)\big] = 0,
\label{eq:app_disp}
\end{equation}
is that obtained by \citet{tam1989instability,tam1990theoretical}. For each azimuthal order, the real solutions $k_z(\omega)$ of equation~\eqref{eq:app_disp} form a discrete family of branches, indexed by $n$ and denoted $A_n$ for $m=0$ and $H_n$ for $m=1$; these are the solid and dashed curves of the dispersion diagram in \S\ref{sec:dispersion} when plotted as $St$ against $k_z D_j$. The upstream-propagating portion of each branch is the segment of negative group velocity $\mathrm{d}\omega/\mathrm{d}k_z < 0$, which defines an allowable frequency band bounded above by the branch turning point and below by the lower limit given next.

\subsection{Allowable-band lower limits}

The lower limit of each branch is set by the loss of radial evanescence in the ambient. Writing the outer radial wavenumber of equation~\eqref{eq:app_mode} as $\kappa_\infty^2 = k_z^2\,(C^2 - 1)$, the mode remains trapped only while $|C| < 1$, that is while its axial phase speed is subsonic with respect to the ambient. At $C = -1$ the outer radial wavenumber vanishes, the $K_m$ solution of equation~\eqref{eq:app_eigfun} ceases to decay, and the branch meets the acoustic-speed line
\begin{equation}
k_z = -\,\frac{\omega}{a_\infty} \quad\Longleftrightarrow\quad k_z D_j = -\,2\pi\,\frac{u_j}{a_\infty}\,St,
\label{eq:app_acline}
\end{equation}
the dashed line of \S\ref{sec:dispersion}. This is a condition on the phase speed alone, and is distinct from the zero-group-velocity turning point that bounds the band from above. Setting $C=-1$ in equation~\eqref{eq:app_disp}, the inside argument becomes $|\xi_-\alpha| = \pi\,(u_j/a_\infty)\,\xi_-^{0}\,St$ with $\xi_-^{0} = |(a_\infty/a_j + M_j)^2 - 1|^{1/2}$. For the axisymmetric modes the ambient loading term of equation~\eqref{eq:app_disp} vanishes in this limit and the relation reduces to $J_1(|\xi_-\alpha|)=0$, the rigid-duct condition $\mathrm{d}\hat p/\mathrm{d}r = 0$ at $r=r_0$, giving the closed-form lower limits
\begin{equation}
St_{\min}^{A_1} = 0, \qquad
St_{\min}^{A_n} = \frac{j_{1,\,n-1}}{\pi\,M_j\,(a_j/a_\infty)\,\big|(a_\infty/a_j + M_j)^2 - 1\big|^{1/2}} \quad (n\ge 2),
\label{eq:app_axisym_min}
\end{equation}
in which $j_{1,\,n-1}$ is the $(n-1)$-th positive zero of $J_1$ \citep{tam1990theoretical}. The helical lower limits have no such elementary form and are the roots of
\begin{equation}
2\,J_1(|\xi_-\alpha|) + \frac{|\xi_-\alpha|}{\big((a_\infty/a_j) + M_j\big)^{2}}\,\big[J_0(|\xi_-\alpha|) - J_2(|\xi_-\alpha|)\big] = 0,
\label{eq:app_helical_min}
\end{equation}
again evaluated on $C=-1$ \citep{tam1990theoretical}. Equations~\eqref{eq:app_eigfun}-\eqref{eq:app_helical_min} are evaluated at the present condition $M_j = 1.56$ and total-temperature ratio unity, for which the ideally expanded jet is colder than the ambient, with $a_j/a_\infty = [1 + \tfrac{1}{2}(\gamma-1)M_j^2]^{-1/2} \approx 0.82$ and $u_j/a_\infty = M_j\,a_j/a_\infty \approx 1.28$, to produce the branches, eigenfunctions and lower-limit corners reported in \S\ref{sec:dispersion}.

\bibliographystyle{jfm}
\bibliography{jfm.bib}

\end{document}